\documentclass[12pt,a4paper,epsf]{article}
\usepackage{graphics}
\usepackage{amssymb,amsmath}
\usepackage[dvips]{lscape,graphicx}
\usepackage{cite}

\newcommand{\ct}{\cite}
\newcommand{\lb}{\label}

\newcommand{\bc}{\begin{center}}
\newcommand{\ec}{\end{center}}
\newcommand{\bd}{\begin{displaymath}}
\newcommand{\ed}{\end{displaymath}}
\newcommand{\be}{\begin{equation}}
\newcommand{\ee}{\end{equation}}
\newcommand{\ba}{\begin{array}}
\newcommand{\ea}{\end{array}}
\newcommand{\bea}{\begin{eqnarray}}
\newcommand{\eea}{\end{eqnarray}}
\newcommand{\bt}{\begin{tabular}}
\newcommand{\et}{\end{tabular}}
\newcommand{\un}{\underline}
\newcommand{\ov}{\overline}
\newcommand{\bp}{\begin{picture}}
\newcommand{\ep}{\end{picture}}
\newcommand{\bfi}{\begin{figure}}
\newcommand{\efi}{\end{figure}}

\begin{document}

\hyphenation{ }

\title{\huge \bf {Mirror World with Broken Mirror Parity, $E_6$ Unification and Cosmology}}
\author{C.R.~Das ${}^{1}$ \footnote{\large\, crdas@pku.edu.cn, crdas@imsc.res.in} ,
L.V.~Laperashvili ${}^{ 2}$ \footnote{\large\, laper@itep.ru} , \\[5mm]
\itshape{${}^{1}$ Center for High Energy Physics, Peking University, Beijing, China}\\[0mm]
\itshape{${}^{2}$ The Institute of Theoretical and Experimental Physics, Moscow, Russia}}

\date{}

\maketitle

\thispagestyle{empty}

\begin{abstract}
In the present paper we have developed a concept of parallel
ordinary (O) and mirror (M) worlds. We have shown that in the case
of a broken mirror parity (MP), the evolutions of fine structure
constants in the O- and M-worlds are not identical. It is assumed
that $E_6$-unification inspired by superstring theory restores the
broken MP at the scale $\sim 10^{18}$ GeV, what unavoidably leads
to the different $E_6$-breakdowns at this scale: $E_6 \to
SO(10)\times U(1)_Z$ - in the O-world, and $E'_6 \to SU(6)'\times
SU(2)'_Z$ - in the M-world. Considering only asymptotically free
theories, we have presented the running of all the inverse gauge
constants $\alpha_i^{-1}$ in the one-loop approximation. Then a
`quintessence' scenario suggested in Refs.~\ct{19,20,20a,21,22,22a} is
discussed for our model of accelerating universe. Such a scenario
is related with an axion (`acceleron') of a new gauge group
$SU(2)'_Z$ which has a coupling constant $g_Z$ extremely growing
at the scale $\Lambda_Z\sim 10^{-3}$ eV.

\end{abstract}

\clearpage\newpage

\pagenumbering{arabic}

\section{Introduction}

In the present paper we consider the concept \ct{1,2} (see also
reviews \ct{3,4}) that there exists in Nature a `mirror' (M) world
-- a hidden mirror sector -- parallel to our ordinary (O) world.
The M-world as a mirror copy of the O-world contains the same
particles and their interactions as our visible world. Observable
elementary particles of our O-world have left-handed (V-A) weak
interactions which violate P-parity. If a hidden mirror M-world
exists, then mirror particles participate in the right-handed
(V+A) weak interactions and have an opposite chirality. Lee and
Yang were first \ct{1} who suggested such a duplication of the
worlds which restores the left-right symmetry of Nature. The term
`Mirror World' was introduced by Kobzarev, Okun and Pomeranchuk in
Ref.~\ct{2}, where they have investigated a lot of
phenomenological implications of such parallel worlds. The
development of this theory is given by
Refs.~\ct{5,5a,5b,5bb,5c,5d,5e,5g,5i,5j,5ja,5jb,5k,5ka,5kb,5kc,5kd,5ke}.

The old idea of the existence of visible and mirror worlds became
very attractive over the last years in connection with a
superstring theory \ct{6,6a,6b,6ba,6bb,6c,6d,6da,7}. Having a theory described
by the product $G\times G'$ of symmetry groups corresponding to
the parallel O- and M-worlds, respectively, it is natural to
associate it with superstring theory described by $E_8\times E'_8$
\ct{6,7}.

Superstring theory is a paramount candidate for the unification of
all fundamental gauge interactions with gravity. Superstrings are
free of gravitational and Yang-Mills anomalies if a gauge group of
symmetry is $SO(32)$ or $E_8\times E_8$. The `heterotic'
superstring theory $E_8\times E'_8$ was suggested as a more
realistic model for unification \ct{6,7}. This ten-dimensional
Yang-Mills theory can undergo spontaneous compactification for
which $E_8$ group is broken to $E_6$ in four-dimensional space.

Among hundreds of papers devoted to the $E_6$-unification we
should like to single out Refs.~\ct{7a,7b,7ba,7bb,7bc,7c,7d,7da,7e,7f,7g}.

If at small distances we have the $E_6$ unification in our
ordinary world, then we can expect to have the same $E_6$
unification in the mirror four-dimensional world assuming a
restoration of the left-right symmetry of Nature at small
distances.

We can consider a minimal symmetry $G_{SM}\times G'_{SM}$, where
$G_{SM} = SU(3)_C\times SU(2)_L\times U(1)_Y$ stands for the
Standard Model (SM) of observable particles: three generations of
quarks and leptons and the Higgs boson, while $G'_{SM} =
SU(3)'_C\times SU(2)'_L\times U(1)'_Y$ is its mirror gauge
counterpart having three generations of mirror quarks and leptons
and the mirror Higgs boson. The M-particles are singlets of
$G_{SM}$ and O-particles are singlets of $G'_{SM}$. These
different O- and M-worlds are coupled only by gravity (or maybe
other very weak interaction).

In this paper all quantities of the mirror world will be marked by
prime ($'$).

A discrete symmetry `MP' of the interchange $G \leftrightarrow G'$
is called `Mirror Parity'. If this parity is conserved, then
particle content of both sectors are identical and described by
the same Lagrangians with the same masses and coupling constants.

The aim of the present paper is to consider a case suggested in
Refs.~\ct{8,9,18}, when a mirror parity MP is not conserved. Then,
as we show, the evolutions of coupling constants in O- and
M-worlds are different. The next assumption that mirror parity is
restored by the $E_6$-unification at the scale $\sim 10^{18}$ GeV
in both O- and M-worlds leads to the significant consequences for
cosmology.

In Section 2 we determine a mirror world and give particle
contents existing in the ordinary and mirror worlds.

Section~3 is devoted to symmetry groups considering only in the
ordinary world. We assume that SM is extended by MSSM (Minimal
Supersymmetric Standard Model). Then its extension by left-right
symmetry leads to the $SO(10)$-unification, which is ended by the
$E_6$-unification at the superGUT scale $M_{SGUT}\sim 10^{18}$
GeV, according to the following chain: 

$$ SU(3)_C\times
SU(2)_L\times U(1)_Y\to \left[SU(3)_C\times SU(2)_L\times
U(1)_Y\right]_{MSSM}$$ $$\to
 SU(3)_C\times SU(2)_L \times SU(2)_R\times U(1)_X\times
 U(1)_Z $$ $$
 \to SU(4)_C\times SU(2)_L \times SU(2)_R\times U(1)_Z$$ $$\to
SO(10)\times U(1)_Z \to E_6\,.$$

All evolutions corresponding to these symmetry groups in the
ordinary world are presented in Figs.~1(a,b) and Figs.~2(a,b) for
supersymmetry breaking scales 1 TeV and 10 TeV, respectively.
Considering only asymptotically free symmetry groups, we have used
the one-loop approximations for these evolutions.

In Section 4 we have considered a mirror world with a broken
mirror parity (MP). We have shown the difference between
evolutions of all fine structure constants in the O- and M-worlds
in the case of broken MP. To get the same $E_6$ unification in
both worlds, we are forced to consider a quite different chain of
the SM$'$ extension in the M-world. We have assumed the following
chain:

$$\left[SU(3)'_C\times SU(2)'_L\times U(1)'_Y\right]_{SM'}\times SU(2)'_Z
$$ $$\to \left[SU(3)'_C\times SU(2)'_L\times SU(2)'_Z\times
U(1)'_Y\right]_{SUSY}$$ $$\to SU(3)'_C\times SU(2)'_L \times
SU(2)'_Z\times U(1)'_X\times
 U(1)'_Z $$ $$\to
SU(4)'_C\times SU(2)'_L\times SU(2)'_Z\times U(1)'_Z$$
$$\to SU(6)'\times SU(2)'_Z \to E'_6\,.$$ 
All the evolutions
corresponding to these symmetry groups in the mirror world are
given in Figs.~3(a,b) and Figs.~4(a,b) (in the one-loop
approximation) for the supersymmetry breaking scale $M'_{SUSY}$.

A comparison of the evolutions considered in both worlds is shown
in Figs.~5(a,b) and Figs.~6(a,b).

All parameters of the evolutions are presented in Table 1.

A new mysterious gauge group $SU(2)'_Z$ is considered in Section
5. We have chosen such a particle content of this group which
leads to the `quintessence' model of our universe. The axion-like
potential is investigated.

A new quintessence scenario in cosmology is developed in Section
6, together with consequences for recent models of dark energy and
dark matter. The problem of cosmological constant also is briefly
discussed in this Section.

\section{Particle content in the ordinary and mirror SM}

In the Standard Model (SM) fermions are represented by Weyl
spinors, however, the left-handed (L) quarks and leptons: $\psi_L
= q_L, l_L$ and right-handed (R) fermions: $\psi_R = q_R, l_R$
transform differently under $SU(2)\times U(1)$ symmetry. A global
lepton charge is $L=1$ for leptons $l_L, l_R$, and a baryon charge
is $B = \frac{1}{3}$ for quarks $q_L, q_R$.

With the same rights we could formulate the SM in terms of
antiparticle fields: $\tilde \psi_R = C\gamma_0\psi_L^*$ and
$\tilde \psi_L = C\gamma_0\psi_R^*$, where $C$ is the charge
conjugation matrix and $\gamma_0$ is the Dirac matrix. These
antiparticles have opposite gauge charges, opposite chirality, $L=
- 1$ for antileptons and $B= - \frac{1}{3}$ for antiquarks.

We can redefine the notion of particles considering L-particles
and R-particles:

\be \rm {L-fermions}: \quad \psi_L\,,\,\, \tilde \psi_L \quad
\rm{and} \quad \rm {\tilde R-fermions}: \quad \tilde \psi_R\,,\,\,
\psi_R\,. \lb{1} \ee 
Including Higgs bosons $\phi$ we have the
following SM content of the O-world: 
$$\rm{L-set}: \quad
(u\,,d\,,e\,,\nu\,,\tilde u\,,\tilde d\,,\tilde e\,,\tilde
N)_L\,,\phi_u\,,\phi_d\,;$$ 
$$ {\rm \tilde R-set}: \quad (\tilde
u\,,\tilde d\,,\tilde e\,,\tilde \nu\,,u\,,d\,,e\,,N)_R\,,
\tilde \phi_u\,,\tilde \phi_d $$ 
\be \quad {\rm with} \quad \tilde \phi_ {u,d} =
\phi^*_{u,d}\,. \lb{2} \ee 
Considering the minimal symmetry $G_{SM}\times G'_{SM}$ we have 
the following particle content in the M-sector:
$$\rm{L'-set}: \quad 
(u'\,,d'\,,e'\,,\nu'\,,\tilde u'\,,\tilde d'\,,\tilde e'\,,\tilde
N')_L\,,\phi'_u\,,\phi'_d\,;$$ 
\be {\rm \tilde R'-set}: \quad
(\tilde u'\,,\tilde d'\,,\tilde e'\,,\tilde \nu'\,,u'\,,
d'\,,e'\,,N')_R\,,\tilde\phi'_u\,,\tilde \phi'_d\,.     \lb{3} \ee

In general, we can consider a supersymmetric theory when $G\times
G'$ contains grand unification groups: $SU(5)\times SU(5)'$,
$SO(10)\times SO(10)'$, $E_6\times E_6'$, etc.

\section{The SM and its extension in the ordinary world}

In the present paper we consider the running of all the gauge
coupling constants in the SM and its extensions which is well
described by the one-loop approximation of the renormalization
group equations (RGEs) from the Electroweak (EW) scale up to the
Planck scale. For simplicity of this investigation and with aim to
demonstrate our idea, we neglect the contributions of higher loops
and Higgs bosons belonging to high representations. Also we do not
pay an attention to the realistic values of some unknown scales
(namely, a supersymmetric scale $M_{SUSY}$, seesaw scale $M_R$,
etc.) which we have used in our numerical calculations. We give
their reasonable values as examples.

For energy scale $\mu \ge M_{ren}$, where $M_{ren}$ is the
renormalization scale, we have the following evolution for the
inverse fine structure constants $\alpha_i^{-1}$ given by RGE in
the one-loop approximation: \be
 \alpha_i^{-1}(\mu) = \alpha_i^{-1}(M_{ren}) + \frac{b_i}{2\pi}t\,,
                                          \lb{4} \ee
where $\alpha_i = g^2_i/4\pi$, $g_i$ are gauge coupling constants
and $t=\ln(\mu/M_{ren})$.

\subsection{Gauge coupling constant evolutions in the SM}

We start with the SM in our ordinary world.

In the SM for energy scale $\mu \ge M_t$ (here $M_t$ is the top
quark (pole) mass) we have the following evolutions (RGEs)
\ct{10,11,12} for the inverse fine structure constants
$\alpha_i^{-1}$ ($i=1,2,3$ correspond to the $U(1)$, $SU(2)_L$
and $SU(3)_C$ groups of the SM), which are revised using updated
experimental results \ct{12} (see also Refs.~\ct{13,14}):

\be
      \alpha_1^{-1}(t) = 58.65 \pm 0.02 - \frac{41}{20\pi}t\,,
                                           \lb{5a}
\ee \be
      \alpha_2^{-1}(t) = 29.95 \pm 0.02 + \frac{19}{12\pi}t\,,
                                            \lb{5b}
\ee \be
      \alpha_3^{-1}(t) = 9.17 \pm 0.20 + \frac{7}{2\pi}t\,.
                                           \lb{5c}
\ee Here $M_{ren}=M_t$ and $t=\ln(\mu/M_t)$. In Eq.~(\ref{5c}) the
value of $\alpha_3^{-1}(M_t) = 9.17$ essentially depends on the
value of $\alpha_3(M_Z)\equiv \alpha_s(M_Z) = 0.118\pm 0.002$ (see
\ct{12}), where $M_Z$ is the mass of $Z$-boson. The value of
$\alpha_3^{-1}(M_t)$ is given by the running of
$\alpha_3^{-1}(\mu)$ from $M_Z$ up to $M_t$, via the Higgs boson
mass $M_H$. We have used the central value of the top quark mass
$M_t\approx 174$ GeV and $M_H = 130\pm 15$ GeV.

Evolutions (\ref{5a})-(\ref{5c}) are shown from $M_t$ up to the
scale $M_{SUSY}$ in Fig.~1(a) and Fig.~2(a), where $x =
\log_{10}\mu$(GeV), $t = x\ln10 - \ln M_t$.

\subsection{Running of gauge coupling constants in the MSSM}

In this Subsection we consider the Minimal Supersymmetric Standard
Model (MSSM) which extends the conventional SM.

MSSM gives the evolutions (\ref{4}) for $\alpha_i^{-1}$
($i=1,2,3$) from the supersymmetric scale $M_{SUSY}$ (here
$M_{ren} = M_{SUSY}$) up to the seesaw scale $M_R$.

Figs.~1(a,b), 2(a,b) present also these evolutions which are given
by the following MSSM slopes $b_i$ \ct{10,11}: \be
           b_1 = - \frac{33}{5} = - 6.6\,, \quad
     b_2 = -1\,, \quad b_3 = 3\,.    \lb{6} \ee
In Figs.~1(a,b), 2(a,b) we have presented examples with the scales
$M_{SUSY} = 1$ and 10 TeV, respectively:

Figs.~1(a,b) are given for SUSY breaking scale $M_{SUSY}= 1$ TeV
and seesaw scale $M_R=1.25\cdot 10^{15}$ GeV; $M_{SGUT}\approx
2.4\cdot 10^{17}$ GeV and $\alpha_{SGUT}^{-1}\approx 26.06$.

Figs.~2(a,b) correspond to SUSY breaking scale $M_{SUSY}= 10$ TeV
and seesaw scale $M_R=2.5\cdot 10^{14}$ GeV; $M_{SGUT}\approx
6.96\cdot 10^{17}$ GeV and $\alpha_{SGUT}^{-1}\approx 27.64$.

\subsection{Left-right symmetry as an extension of the MSSM}

At the seesaw scale $M_R$ the heavy right-handed neutrinos appear.
We assume that the following supersymmetric left-right symmetry
\ct{15,15a,15aa,15b} originates at the seesaw scale: \be
 SU(3)_C\times SU(2)_L\times
SU(2)_R\times U(1)_X\times U(1)_Z\,. \lb{7} \ee Here we see
additional groups $SU(2)_R$ and $U(1)_{(B-L)}$ originated at the
scale $M_R$. The group $U(1)_{(B-L)}$ is mixed with the gauge
group $U(1)_Y$ leading to the product $U(1)_X\times U(1)_Z$ of the
two $U(1)$ groups with quantum numbers $X$ and $Z$ linearly
combined into the weak hypercharge $Y$ \ct{16}: \be
         \frac{Y}{2} = \frac{1}{5}(X - Z)\,. \lb{8} \ee
Considering the running (\ref{4}) for the supersymmetric group
(\ref{7}) in the region $\mu\ge M_R$ we have $M_{ren} = M_R$ and
the following slopes \ct{10,11}:
\be
       b_X = b_1 = - 6.6\,, \quad b_Z = - 9\,, \quad b_3 = 3\,.
                                          \lb{9} \ee
Also  the running of $SU(2)_L\times SU(2)_R$ in the same region of
$\mu$ is given by the slope:
\be
                 b_{22} = - 2\,,   \lb{10} \ee
and we have the following evolution: \be
 \alpha_{22}^{-1}(\mu) = \alpha_{22}^{-1}(M_R) + 
\frac{1}{\pi}\ln\frac{\mu}{M_R}\,,
                                          \lb{10a} \ee
where
\be
         \alpha_{22}^{-1}(M_R) = \alpha_2^{-1}(M_R)\,.
                                                       \lb{10b} \ee
The next step is an assumption that the group $SU(4)_C\times
SU(2)_L\times SU(2)_R$ by Pati and Salam \ct{15} originates at the
scale $M_4$ giving the following extension of the group (\ref{7}):
\be
 SU(3)_C\times SU(2)_L\times
SU(2)_R\times U(1)_X\times U(1)_Z \to  SU(4)_C\times SU(2)_L\times
SU(2)_R\times U(1)_Z\,.  \lb{11} \ee
The scale $M_4$ is given by the intersection of $SU(3)_C$ with
$U(1)_X$:
\be
        \alpha_3^{-1}(M_4) = \alpha_X^{-1}(M_4)\,.   \lb{12} \ee
In the MSSM we have the following equation for the slope $b_N$ of
the $SU(N)$ group (see \ct{10,11} and \ct{7f}):
 \be
     b_N = 3N - N_f - \frac 12 N_{vector} - N\cdot N_{adjoint} -
     ...\,,                          \lb{13} \ee
where $N_f$ is a number of flavors, $N_{vector}$ is a number of
scalar Higgs fields in the fundamental representation and
$N_{adjoint}$ is a number of Higgses in adjoint representation.
Considering only the minimal content of scalar Higgs fields, e.g.
quartets $4+\bar 4$, we have $N_{vector} = 2$ and obtain from
(\ref{13}) the following slope for the running of
$\alpha_4^{-1}(\mu)$: \be
       b_4 = 3\cdot 4 - 6 - 1 = 5\,.  \lb{14}     \ee
Now the evolution (\ref{4}) with $M_{ren} = M_4$ gives: \be
 \alpha_4^{-1}(\mu) = \alpha_4^{-1}(M_4) +
 \frac{5}{2\pi}\ln\frac{\mu}{M_4}\,.
                                          \lb{14a} \ee
This is the running for the symmetry group $SU(4)$.

\subsection{From $SO(10)$ to the $E_6$-unification in the ordinary world }

The intersection of $\alpha_4^{-1}(\mu)$ with the running of
$\alpha_{22}^{-1}(\mu)$ leads to the scale $M_{GUT}$ of the
$SO(10)$-unification: \be
        SU(4)_C\times SU(2)_L \times SU(2)_R \to SO(10)\,,
                                             \lb{15}     \ee
and we obtain the value of $M_{GUT}$ from the relation:
\be
 \alpha_4^{-1}(M_{GUT}) = \alpha_{22}^{-1}(M_{GUT})\,.   \lb{16} \ee
Then we deal with the running (\ref{4}) for the $SO(10)$ inverse
gauge constant $\alpha_{10}^{-1}(\mu)$, which runs from the scale
$M_{GUT}$ up to the scale $M_{SGUT}$ of the super-unification
$E_6$: \be
     SO(10)\times U(1)_Z \to E_6\,.    \lb{17}     \ee
The slope of this running is $b_{10}$.

In general, for the $SO(N)$ group we have the following slope
\ct{7f,10,11}:

 \be
     b_N^{SO(N)} = \frac 32 (N-2) - N_f - \frac 12 N_{vector} - 
\frac 12 (N-2)\cdot N_{adjoint} -
     ...\,.                          \lb{18} \ee
Calculating the $SO(10)$-slope we must consider not only vectorial
Higgs fields $N_{vector} = 2$, but also $N_{adjoint} = 1$, because
the appearance of right-handed particles is impossible without
adjoint Higgs field (see explanation in Ref.~\ct{14}). As a
result, we obtain from Eq.~(\ref{18}) the following
$SO(10)$-slope:
\be
         b_{10} = 12 - 6 - 1 - 4 = 1\,.
                                 \lb{19} \ee
Then we have the following running of $\alpha_{10}^{-1}(\mu)$: \be
 \alpha_{10}^{-1}(\mu)
 = \alpha_{10}^{-1}(M_{GUT}) +
 \frac{1}{2\pi}\ln \frac{\mu}{M_{GUT}}\,,
                                              \lb{20} \ee
which is valid up to the superGUT scale $M_{SGUT}$ of the
$E_6$-unification.

Finally, as a result of our investigation, one can envision the
following symmetry breaking chain in the ordinary world:
$$
     E_6 \to SO(10)\times U(1)_Z\to SU(4)_C\times 
SU(2)_L \times SU(2)_R\times U(1)_Z\to $$
     $$SU(3)_C\times SU(2)_L \times SU(2)_R\times U(1)_X\times U(1)_Z
     \to  SU(3)_C\times SU(2)_L\times U(1)_Y\,. $$
All evolutions of the corresponding inverse fine structure
constants are given in Figs.~1(a,b) and 2(a,b).

\section{Mirror world with broken mirror parity and mirror scales}

In this Section, as in Refs.~\ct{8,9,18} (see also \ct{3,4}), our
main assumption is the principle: ``The only good parity... is a
broken parity", what means that in general case the mirror parity
MP is not conserved in Nature. However, at the very small
distances the mirror parity is restored and super-unifications
$E_6$ and $E_6'$ (inspired by superstring theory) are identical
having the same  $M_{SGUT}\sim 10^{17}$ or $\sim 10^{18}$ GeV. By
this reason, the superGUT scale $M_{SGUT}$ may be fixed by the
intersection of the evolutions of gauge coupling constants in both
-- mirror and ordinary -- worlds, which were not identical from
the beginning.

Now it is very interesting to discuss what particle physics exists
in the mirror world when the mirror parity MP is spontaneously
broken.

If O- and M-sectors are described by the minimal SM with the Higgs
doublets $\phi$ and $\phi'$, respectively,  then we can consider
the Higgs potentials:
\be
       U = -\mu^2 \phi^+\phi + \frac{\lambda}{4} (\phi^+\phi)^2\,,
                \lb{21a} \ee
and \be
       U' = -\mu'^2 \phi'^+\phi' + \frac{\lambda'}{4}
       (\phi'^+\phi')^2\,.
                                                \lb{21} \ee
In the case of non-conserved MP the VEVs of $\phi$ and $\phi'$ are
not equal:

\be
       v = \frac {2\mu}{\lambda} \neq  v' = \frac
       {2\mu'}{\lambda'}\,.
                                                  \lb{22} \ee
Following Refs.~\ct{8,9,18}, we assume that $v' \gg v$ and
introduce the parameter characterizing the violation of MP: \be
        \zeta = \frac{v'}{v} \gg 1\,.   \lb{23} \ee
As far as Yukawa couplings have the same values in both worlds,
the masses of the SM fermions and massive bosons in the mirror
world are scaled up by the factor $\zeta$:
$$
               m'_{q',l'} = \zeta m_{q,l}\,,  $$
\be
                 M'_{W',Z',\phi'} = \zeta M_{W,Z,\phi}\,,
                                      \lb{24} \ee
but photons and gluons remain massless in both worlds.

Let us consider now the following expressions:
$$
    \alpha_i^{-1}(\mu) = \frac{b_i}{2\pi}\ln
    \frac{\mu}{\Lambda_i}\,,$$
--- in the O-world, and
\be {\alpha'}_i^{-1}(\mu) = \frac{b'_i}{2\pi}\ln
    \frac{\mu}{\Lambda'_i}
                               \lb{25}  \ee
--- in the M-world.

A big difference between $v$ and $v'$ will not cause a big
difference between scales $\Lambda_i$ and $\Lambda'_i$ (see
\ct{4,9}): \be
            \Lambda'_i = \xi \Lambda_i   \quad
\rm{with}\quad \xi > 1\,.  \lb{26}  \ee The values of $\zeta $ and
$\xi$ were estimated in Refs.~\ct{4,8,9,18}: \be \zeta \approx 30
\quad \rm{and} \quad
                         \xi\approx 1.5\,, \lb{26a} \ee
as results of astrophysical implications of the mirror world with
broken mirror parity. But it is possible to have $\zeta $ in the
region: \be  10 \le \zeta \le 100\,.  \lb{26b} \ee As for the
neutrino masses, the same authors have shown that the theory with
broken mirror parity leads to the following relations:
$$
      m'_{\nu} = \zeta^2 m_{\nu}\,, $$ and
\be
      M'_{\nu} = \zeta^2 M_{\nu}\,,   \lb{27}  \ee
where $m_{\nu}$ are light left-handed and $M_{\nu}$ are heavy
right-handed neutrino masses in the O-world, and
$m'_{\nu},M'_{\nu}$ are the corresponding neutrino masses in the
M-world. These relations are valid for each of three generations.

\subsection{Broken mirror parity and the running of gauge coupling 
constants in the mirror SM}

We assume that M-world is not P- and CP- invariant and differs
with O-world.

Considering the mirror SM given by symmetry group $G'_{SM} =
SU(3)'_C\times SU(2)'_L\times U(1)'_Y$, we deal with the following
one-loop approximation RGE for the running of inverse fine
structure constants ${\alpha'}_i^{-1}(\mu)$ ($i=1,2,3$ correspond
to the $U(1)'$, $SU(2)'_L$ and $SU(3)'_C$ groups of the mirror SM
with broken MP): \be
 {\alpha'}_i^{-1}(\mu) = {\alpha'}_i^{-1}(M'_{ren}) + \
\frac{b'_i}{2\pi}t'\,,
                                          \lb{4'} \ee
where $M'_{ren} = \zeta M_{ren}$ is the renormalization scale in
the mirror world and $t'=\ln(\mu/M'_{ren})$.

In the M-world we have scales $\Lambda'_i$ which are different
with $\Lambda_i$ (they are given by Eq.~(\ref{26})), but O- and
M-slopes are identical:
\be
         b'_i = b_i\,.     \lb{28}  \ee

Then in the SM of the M-sector we have the following evolutions:
\be
 {(\alpha')}_i^{-1}(\mu) = {(\alpha')}_i^{-1}(M'_t) + 
\frac{b_i}{2\pi}t' = \frac{b_i}{2\pi}\ln \frac{\mu}{\Lambda'_i}\,,
                                                        \lb{29} \ee
where
\be
{(\alpha')}_i^{-1}(M_t) = \alpha_i^{-1}(M_t) - \frac{b_i}{2\pi}\ln \xi\,,
                                                \lb{30} \ee
or
\be
           {(\alpha')}_i^{-1}(M'_t) = \alpha_i^{-1}(M_t)\,. \lb{31} \ee
Finally, we obtain the following SM running of gauge coupling
constants in the mirror world:

\begin{enumerate}
\item[(i)] \be
    {(\alpha')}_1^{-1}(\mu) = 58.65 \pm 0.02
                      - \frac{41}{20\pi}t'\,,
                     \lb{32a}  \ee

\item[(ii)] \be
    {(\alpha')}_2^{-1}(\mu) = 29.95 \pm 0.02 +
    \frac{19}{12\pi}t'\,,
                     \lb{32b}  \ee

\item[(iii)] \be
    {(\alpha')}_3^{-1}(\mu) = 9.17 \pm 0.20 + \frac{7}{2\pi}t'\,,
                                 \lb{32c}  \ee
\end{enumerate}
where $t' = \ln{(\mu/M'_t)}$. According to Eq.~(\ref{24}), the
pole mass of the mirror top quark is $M'_t =\zeta M_t$.

\subsection{Mirror MSSM and a seesaw scale in the mirror world}

If the Minimal Supersymmetric Standard Model (MSSM) extends the
mirror SM in the mirror world, then we can assume that mirror
sparticle masses $\tilde m'$ obey the relation analogous to
Eq.~(\ref{24}): \be
         \tilde m' = \zeta \tilde m\,.  \lb{33} \ee
This relation leads to the assumption that the mirror
SUSY-breaking scale is larger than $M_{SUSY}$:
\be
      M'_{SUSY} = \zeta M_{SUSY}\,.     \lb{34} \ee
The mirror MSSM gives the evolutions (\ref{4'}) for
${\alpha'}_i^{-1}(\mu)$ ($i=1,2,3$) from the supersymmetric scale
$M'_{SUSY}$ (here $M'_{ren} = M'_{SUSY}$) up to the GUT scale
$M'_{GUT}$.

Here it is worth the reader's attention to observe that if heavy
right-handed neutrino masses are given by Eq.~(\ref{27}), then a
mirror seesaw scale $M'_R$ obeys the following relation: \be
          M'_R = \zeta^2 M_R\,.  \lb{35} \ee
According to the estimate (\ref{26a}) given by
Refs.~\ct{4,8,9,18}, we have:
\be
          M'_R \sim 10^{3} M_R\,. \lb{36} \ee
Now if $M_R \sim 10^{14}$ GeV, then $M'_R \sim 10^{17}$ GeV, and a
seesaw scale is close to the superGUT scale of the
$E_6$-unification. This means that mirror heavy right-handed
neutrinos appear at the scale $\sim 10^{17}$ GeV.

Figs.~3(a), 4(a) present the mirror MSSM evolutions of
${\alpha'}_i^{-1}(\mu)$ ($i=1,2,3$), where slopes $b_i$ are given
by the same Eq.~(\ref{6}) as in the O-world \ct{10,11}. In
Figs.~3(a,b) we have presented an example of the mirror MSSM
evolution with the scale $M'_{SUSY} = 10$ TeV, what corresponds to
$M_{SUSY} = 1$ TeV and $\zeta = 10$. But in Figs.~4(a,b) we have
shown an example of the mirror MSSM evolution with the scale
$M'_{SUSY} = 300$ TeV, what corresponds to $M_{SUSY} = 10$ TeV and
$\zeta = 30$.

\subsection{From $SU(6)$ to the $E_6$-unification in the mirror world}

Let us consider now the extension of the MSSM in the mirror world.

The first step of such an extension is: \be [SU(3)'_C\times
SU(2)'_L\times U(1)'_Y]_{MSSM} \to  SU(4)'_C\times SU(2)'_L\,.
\lb{11} \ee Assuming that the supersymmetric group $SU(4)'_C\times
SU(2)'_L$ originates at the scale $M'_4$, we find
the intersection of $SU(3)'_C$ with $U(1)'_Y$:
\be
   {\alpha'}_3^{-1}(M'_4) = \alpha_Y^{-1}(M'_4)\,.  \lb{37} \ee
The gauge symmetry group $SU(4)'_C$ starts from the scale $M'_4$,
and we have the following evolution: \be
 {(\alpha')}_4^{-1}(\mu) = {(\alpha')}_4^{-1}(M'_4) +
 \frac{5}{2\pi}\ln \frac{\mu}{M'_4}\,,
                                              \lb{38} \ee
which runs up to the intersection with the running
${(\alpha')}_2^{-1}(\mu)$ for the supersymmetric group $SU(2)'_L$
having $b_2=-1$. The point of this intersection is the scale
$M'_{GUT}$, which is given by the following relation:
\be
 {(\alpha')}_4^{-1}(M'_{GUT}) = {(\alpha')}_2^{-1}(M'_{GUT})\,.   
\lb{39} \ee
At the mirror GUTscale $M'_{GUT}$ we obtain the
$SU(6)'$-unification: \be
 SU(4)'_C\times SU(2)'_L\times U(1)'_Z  \to SU(6)'\,. \lb{40} \ee
We see that $U(1)'_Z$ also meets $SU(4)'_C$ and $SU(2)'_L$ at the
same GUTscale.

For $\mu\ge M'_{GUT}$ we must consider the running of
${(\alpha')}_6^{-1}(\mu)$ up to the superGUT scale  $M'_{SGUT} =
M'_{E6}$: \be {(\alpha')}_6^{-1}(\mu) =
{(\alpha')}_6^{-1}(M'_{GUT}) +
 \frac{b_6}{2\pi}\ln \frac{\mu}{M'_{GUT}} =
{(\alpha')}_6^{-1}(M'_{GUT}) +
 \frac{11}{2\pi}\ln \frac{\mu}{M'_{GUT}}\,,
                                              \lb{41} \ee
where we have used the result
\be
    b_6 = 11\,,   \lb{42} \ee
obtained from Eq.~(\ref{13}) for $N=6$, $N_f=6$, $N_{vector}=2$.
Here we assumed the existence of only minimal number of the Higgs
fields, namely $h + \bar h$, belonging to the fundamental
representation \un 6 of the $SU(6)'$ group.

Now it is obvious that we must find some unknown in the O-world
symmetry group $SU(2)'_Z$, which must help us to reach the
desirable $E'_6$-unification at the superGUT scale $M'_{SGUT}$:
\be SU(6)'\times SU(2)'_Z \to E'_6\,.   \lb{43} \ee In the present
investigation we assume that $E_6$-unifications restore the
breakdown of the mirror parity MP and $M'_{SGUT} = M_{SGUT} =
M_{E6}$. Then the scale $M_{SGUT}$ of the $E_6\times
E'_6$-unification is given by the following intersection: \be
    \alpha_{10}^{-1}(M_{SGUT}) =  {(\alpha')}_6^{-1}(M_{SGUT})\,.
                                              \lb{44} \ee
Finally, one can envision the following symmetry breaking chain in
the M-world:
$$
E'_6 \to SU(6)'\times SU(2)'_Z \to SU(4)'_C\times SU(2)'_L\times
SU(2)'_Z\times U(1)'_Z$$ $$\to SU(3)'_C\times SU(2)'_L\times
SU(2)'_Z\times U(1)'_X\times U(1)'_Z$$ $$
\to SU(3)'_C\times
SU(2)'_L\times SU(2)'_Z\times U(1)'_Y\,.
$$
Here it is quite necessary to understand if there  exists  the
group $SU(2)'_Z$ in the mirror world. What it could be?

\section{The mirror $SU(2)'_Z$ and cosmological constant}

A candidate for such a gauge group $SU(2)'_Z$, which could be
unified with $SU(6)'$ in the mirror world (see Eq.~(\ref{43})),
was suggested in Refs.~\ct{19,20,20a,21,22,22a}.

We shall consider now the possibilities presented by Ref.~\ct{19}
and recently by Refs.~\ct{20,20a,21,22,22a} (see also \ct{32,31}), where a
new gauge group $SU(2)'_Z$ was aimed to introduce a new dynamical
scale $\Lambda_Z\sim 10^{-3}$ eV, which is consistent with present
measurements of cosmological constant \ct{24,25,25a,26,27,27a,28}: a total
vacuum energy density of our universe (named cosmological
constant) is equal to the following value: \be \rho_{vac}\approx
(3\times 10^{-3}\,\,\rm{eV})^4 \lb{45} \ee (see also Ref.~\ct{23a}
for details).

In the model \ct{20,20a,21,22,22a} an axion-like scalar field $a_Z$
(`acceleron') is related with $SU(2)'_Z$, which exists in the
mirror world. The effective potential considered as a function of
the norm of this scalar field has a minimum (so called `false'
vacuum) just with vacuum energy density (\ref{45}).

In the present investigation we assume that the analogous gauge
group $SU(2)'_Z$ leads to our $E_{6}'$-unification in the mirror
world. Then a low energy symmetry group of the M-sector is: \be G'
= SU(3)'_C\times SU(2)'_L\times SU(2)'_Z\times U(1)'_Y =
G'_{SM}\times SU(2)'_Z\,.
                                            \lb{45a} \ee
A new asymptotically free gauge group $SU(2)'_Z$ gives the running
of its inverse fine structure constant
${(\alpha')}_{2Z}^{-1}(\mu)$, which has to grow from the extremely
low energy scale $\Lambda_Z\sim 10^{-3}$ eV up to the
supersymmetric scale $M_{SUSY}$ and then continue to run (in our
model -it does not change, see Figs.~3(a) and 4(a))\, up to the
superGUT scale $M'_{SGUT} = M_{E6}\sim 10^{18}$ GeV. But this
running essentially depends on the particle content of $SU(2)'_Z$.

\subsection{Particle content of $SU(2)'_Z$ gauge group}

For the case $\zeta = 30$  we have obtained the content of
$SU(2)'_Z$ particles, which is not identical with the one given by
Refs.~\ct{19,20,20a,21,22,22a}. The reason of our choice was to obtain the
correct running of ${(\alpha')}_{2Z}^{-1}(\mu)$, which leads to
the scale $\Lambda_Z\sim 10^{-3}$ eV  and simultaneously is
consistent with our specific description of the running of all the
inverse gauge coupling constants in the ordinary and mirror worlds
with broken mirror parity.

Considering the evolutions (\ref{4'}), we have used the following
equations for slopes $b_{2Z}$ \ct{10,11} (see also \ct{7f}): \be
       b_{2Z} = \frac{22}{3} - \frac{4}{3}N_g - \frac{8}{3}N_F -
       \frac{1}{6}N_{vec} - \frac{2}{3}N_{adj} - ...   \lb{46}
\ee
--- for non-supersymmetric $SU(2)$, and
\be
      b_{2Z}^{SUSY} = 6 - 2N_g - 4N_F - \frac{1}{2}N_{vec}^{SUSY} -
      2N_{adj} - ...           \lb{47}
\ee
--- for supersymmetric $SU(2)$.

In Eqs.~(\ref{46}) and (\ref{47}) we have: the number of fermion
doublets $N_g$, the number of fermion triplets $N_F$, the number
of scalar doublets $N_{vec}$ with $N_{vec}^{SUSY} = 2N_{vec}$ and
the number of triplet scalars $N_{adj}$.

Only the following slopes are consistent with our description of
the O- and M- sectors, which can give the correct scale
$\Lambda_Z\sim 10^{-3}$ eV:
 \be
b_{2Z} =\frac{13}{3}\approx 4.33\quad {\rm and}\quad  b_{2Z}^{SUSY}= 0\,.
                                               \lb{48} \ee
Then the particle content of $SU(2)'_Z$ is as follows:

\begin{itemize}

\item two doublets of fermions $\psi^{(Z)}_i$ and two
doublets of `messenger' scalar fields $\phi^{(Z)}_i$ with $i =
1,2$ (here $N_g=2$, $N_F=0$, $N_{vec}=2$ and $N_{vec}^{SUSY}=4$),

or

\item one triplet of fermions $\psi^{(Z)}_f$ with
$f=1,2,3$, which are singlets under the SM, and two doublets of
`messenger' scalar fields $\phi^{(Z)}_i$ with $i = 1,2$ (in this
case $N_g=0$, $N_F=1$, $N_{vec}=2$ and $N_{vec}^{SUSY}=4$).

\item We also consider a complex singlet scalar field
$\varphi_Z$: \be \varphi_Z = (1,1,1,0) \lb{48a} \ee under the
symmetry group $G' = SU(3)'_C\times SU(2)'_L\times SU(2)'_Z\times
U(1)'_Y$.

\end{itemize}

The so called `messenger' fields $\phi^{(Z)}$ carry quantum
numbers of both the SM$'$ and $SU(2)'_Z$ groups. They have Yukawa
couplings with SM$'$ leptons and fermions $\psi^{(Z)}$.

All the SM$'$ particles are assumed to be singlets under $SU(2)'_Z$.

As a result, we obtain the following evolutions:

\begin{enumerate}
\item[(i)] for the region $\mu \le M'_{SUSY}$:
 \be
 {\alpha'}_{2Z}^{-1}(\mu) = {\alpha'}_{2Z}^{-1}(M'_t) +
 \frac{b_{2Z}}{2\pi}\ln\frac{\mu}{M'_t}\approx
 \frac{b_{2Z}}{2\pi}\ln\frac{\mu}{\Lambda_Z}\,,
                                          \lb{49} \ee

\item[(ii)] and for the region $M'_{SUSY}\le \mu \le M'_{SGUT}$: \be
{\alpha'}_{2Z}^{-1}(\mu) = {\alpha'}_{2Z}^{-1}(M'_{SUSY}) +
 \frac{b_{2Z}^{SUSY}}{2\pi}\ln\frac{\mu}{M'_{SUSY}}\,.
                                       \lb{50} \ee

\end{enumerate}

Also we have the following relation:
\be
        {\alpha'}_{2Z}^{-1}(M'_{SGUT}=M_{SGUT}) = \alpha^{-1}_{E6}\,.
                                  \lb{51} \ee

According to Eq.~(\ref{48}), we have:

\begin{enumerate}

\item[(i)] for the region $\mu \le M'_{SUSY}$: \be
 {\alpha'}_{2Z}^{-1}(\mu) = {\alpha'}_{2Z}^{-1}(M'_t) +
 \frac{13}{6\pi}\ln\frac{\mu}{M'_t}\approx
 \frac{13}{6\pi}\ln\frac{\mu}{\Lambda_Z}\,,
                                          \lb{49a} \ee

\item[(ii)] and in the region $M'_{SUSY}\le \mu \le M'_{SGUT}$ the
evolution ${\alpha'}_{2Z}^{-1}(\mu)$ is unchanged: \be
{\alpha'}_{2Z}^{-1}(\mu) = {\alpha'}_{2Z}^{-1}(M_{SGUT})\,,
                                       \lb{50a} \ee
or \be
      {\alpha'}_{2Z}^{-1}(M'_{SUSY}) =  {\alpha'}_{2Z}^{-1}(M_{SGUT}) 
= \alpha^{-1}_{E6}\,.
                                  \lb{51a} \ee

\end{enumerate}

In Figs.~3(a,b) and 4(a,b) we have shown all evolutions in the
mirror world:

Figs.~3(a,b) are given for SUSY breaking scale $M'_{SUSY}= 10$ TeV
and mirror seesaw scale $M'_R=1.44\cdot 10^{17}$ GeV; $\zeta =
10$; $M_{SGUT}\approx 2.4\cdot 10^{17}$ GeV and
$\alpha_{SGUT}^{-1}\approx 26.06$.

Figs.~4(a,b) correspond to SUSY breaking scale $M'_{SUSY}= 300$
TeV and mirror seesaw scale $M'_R=2.25\cdot 10^{17}$ GeV; $\zeta =
30$; $M_{SGUT}\approx 6.96\cdot 10^{17}$ GeV and
$\alpha_{SGUT}^{-1}\approx 27.64$.

The total pictures of the evolutions in the O- and M-worlds
simultaneously are presented in Figs.~5(a,b) and 6(a,b) for the
cases $M_{SUSY} = 1$ and 10 TeV, $M_R = 1.25\cdot
10^{15}$ GeV and $M_R = 2.5\cdot 10^{14}$ GeV, $\zeta = 10$ and
$\zeta = 30$, respectively. It is obvious that respectively
$M'_{SUSY} = 10$ and 300 TeV; $M'_R=1.44\cdot
10^{17}$ GeV and $M'_R=2.25\cdot 10^{17}$ GeV. Here
$M_{SGUT}\approx 2.4\cdot 10^{17}$ GeV and
$\alpha_{SGUT}^{-1}\approx 26.06$ -- for Figs.~5(a,b), and
$M_{SGUT}\approx 6.96\cdot 10^{17}$ GeV and
$\alpha_{SGUT}^{-1}\approx 27.64$ -- for Figs.~6(a,b).

All parameters  of these evolutions are presented in Table 1.

\begin{table}[p]\caption{}
\begin{tabular}{|c|c|c|c|c|c|c|c|}\hline
Sl. & $M_{SUSY}$ & $M_R$               & $\zeta$ & $M'_{SUSY}$ & 
$M'_{R}$            & $M_{SGUT}$          & $\alpha_{SGUT}^{-1}$\\
No. & in TeV     & in GeV              &         & in TeV      & 
in GeV              & in GeV              &                     \\\hline
    &            &                     &         &             &  
                   &                     &                     \\
1   & 1          & $1.25\cdot 10^{15}$ &  10     & 10          & 
$1.44\cdot 10^{17}$ & $2.4\cdot 10^{17}$  & 26.06               \\\hline
    &            &                     &         &             & 
                    &                     &                     \\
2   & 10         & $2.5\cdot 10^{14}$  &  30     & 300         & 
$2.25\cdot 10^{17}$ & $6.96\cdot 10^{17}$ & 27.64               \\\hline
\end{tabular}
\end{table}

\subsection{The axion potential}

As it was shown in Refs.~\ct{20,20a,21,22,22a}, the Lagrangian
corresponding to the group of symmetry (\ref{45a}) exhibits a
$U(1)_A^{(Z)}$ global symmetry.

A singlet complex scalar field $\varphi_Z$ was introduced in
theory with aim to reproduce a model of Peccei-Quinn (PQ)
(well-known in QCD) \ct{23}. Then the potential:
 \be
     V = \frac{\lambda}{4}(\varphi^+_Z\varphi_Z - v_Z^2)^2
                                         \lb{52} \ee
gives the VEV for $\varphi_Z$: \be \langle \varphi_Z\rangle = v_Z\,.
                   \lb{53} \ee
Representing the field $\varphi_Z$ as follows:
 \be
      \varphi_Z = v_Z\exp(ia_Z/v_Z) + \sigma_Z\,,
                             \lb{54} \ee
we have: \be \langle a_Z\rangle = \langle \sigma_Z\rangle = 0\,.
                       \lb{55} \ee
A boson $a_Z$ (the imaginary part of a singlet scalar field
$\varphi_Z$) is an axion and could be a massless Nambu-Goldstone
(NG) boson if the $U(1)_A^{(Z)}$ symmetry is not spontaneously
broken. However, the spontaneous breakdown of the global
$U(1)_A^{(Z)}$ by $SU(2)'_Z$ instantons gives masses to fermions
$\psi^{(Z)}$ and inverts $a_Z$ into a pseudo Nambu-Goldstone boson
(PNGB) with a mass squared \ct{20,20a,21,22,22a}: \be m_a^2\sim
\Lambda_Z^3/v_Z\sim 10^{-30}\,\,{\rm{GeV}}^2\,.
                               \lb{55a} \ee
Then the field $\varphi_Z$ becomes:
 \be
      \varphi_Z(x) = \exp(ia_Z/v_Z)(v_Z + \sigma(x))\approx
           v_Z + \sigma(x) +ia_Z(x)\,. \lb{56} \ee
Here the field $\sigma $ is an inflaton.

The axion potential is given by the PQ model \ct{23} and has (for
small $a_Z$) the following expression: \be
    V_{axion}\approx  \frac{\lambda}{4}(\varphi^+_Z\varphi_Z - v_Z^2)^2 + K
    |\varphi|\cos(a_Z/v_z)\,,               \lb{57} \ee
where $K$ is a positive constant.

It is well-known that this potential exhibits two degenerate
minima at $\langle a_Z\rangle = 0$ and at $\langle a_Z\rangle = 
2\pi v_Z$ with the potential
barrier existing between them (see Fig.~7).

The minimum of the above-mentioned potential at $\langle a_Z\rangle = 0$
corresponds to the `true' vacuum, while the minimum at $\langle a_Z\rangle =
2\pi v_Z$ is called the `false' vacuum. Such properties of the
present axion lead to the `quintessence' model of our expanding
universe. By this reason, the axion $a_Z$ could be called an
`acceleron'.

\section{Quintessence model of the universe with broken mirror parity}

Recent models of the Dark Energy (DE) and Dark Matter (DM) are
based on measurements in contemporary cosmology
\ct{24,25,25a,26,27,27a,28}. Supernovae observations at redshifts ($1.25
\le z \le 1.7$) by the Supernovae Legacy Survey (SNLS), cosmic
microwave background (CMB), cluster data and baryon acoustic
oscillations by the Sloan Digital Sky Survey (SDSS) fit the
equation of state for DE: $w = p/\rho$ with constant $w$, which is
given by Ref.~\ct{28}: $w = -1.023 \pm 0.090 \pm 0.054$. The value
$w\approx - 1$ is consistent with the present quintessence model
of accelerating universe \ct{30,30a,30b,30c,30d,30da} (see also reviews
\ct{31a,31b} and recent Ref.~\ct{31c}), dominated by a tiny
cosmological constant and Cold Dark Matter (CDM) -- this is a so
called $\Lambda CDM$ scenario \ct{32}.

Here we present the quintessence scenario, which was developed in
Refs.~\ct{20,20a,21,22,22a} in connection with the existence of a new
gauge group $SU(2)'_Z$. We have an analogous situation in our
model, inspired by the broken mirror parity, although the details
of our theory are different. However, in both cases the
cosmological implications of suggested scenarios lead to the
$\Lambda CDM$ model.

\subsection{Dark energy and cosmological constant}

For the ratios of densities $\Omega_X =\rho_X/\rho_c$, where
$\rho_c$ is the critical energy density, cosmological measurements
gave: $\Omega_B\sim $4\% for baryons (visible and dark),
$\Omega_{DM}\sim $23\% for non-baryonic DM, and $\Omega_{DE}\sim
$73\% for the mysterious DE, which is responsible for the
acceleration of our universe.

In Section 5 we have considered that a cosmological constant ($CC$)
is given by the value  $CC = \rho_{vac}\approx (3\times
10^{-3}\,\,\rm{eV})^4$ \ct{24,25,25a,26,27,27a,28}. The main assumption of
Refs.~\ct{22,22a} is the following idea: the universe is trapped in the
false vacuum with $CC$ given by (\ref{45}), but at the end it must
decay into the true vacuum with vanishing $CC$. The true Electroweak
vacuum would have its vacuum energy density $CC = \rho_{vac} = 0$.
Such a scenario exists in the model with Multiple Point Principle
(MPP) \ct{33,33aa,33a,33ab,34} (see also reviews \ct{35,35a} and references
there). A non-zero (nevertheless tiny) $CC$ would be associated only
with a false vacuum. Why $CC$ is zero in a true vacuum? It is a
non-trivial problem, but Refs.~\ct{36,37} try to give a solution
of this problem.

It follows from the last Section 5 that the potential $V(a_Z)$
determines the origin of DE. As it was shown in Ref.~\ct{21}, we
see that when the temperature of the universe $T$ is high: $T
\gg \Lambda_Z$, then the axion potential is flat because the
effects of the $SU(2)'_Z$ instantons are negligible for such
temperatures.

When the temperature begins to decrease, the universe gets trapped
in the false vacuum. At $T\sim \Lambda_Z$ the true vacuum at
$\langle a_Z\rangle = 0$ has zero energy density (cosmological constant $CC$),
and the barrier between vacua is higher. The difference in energy
density between the true and false vacua is now $\Lambda_Z^4$. The
universe is still trapped in the false vacuum with $CC =
\rho_{vac} = \Lambda_Z^4$.

The first order phase transition to the true vacuum is provoked by
the bubble nucleation. In fact, the universe lives in the false
vacuum for a very long time.  When the universe is trapped into
the false vacuum at $\langle a_Z\rangle = 2\pi v_Z$, the deceleration stops and
acceleration begins at $\ddot a_Z = 0$, then ${\dot a}^2_Z =0$ and
$w(a_Z)=-1$. The total energy density of the universe is dominated
by the energy density of the false vacuum, and our universe
undergoes an exponential expansion.

The universe trends to get the true vacuum, which has zero $CC$, but
will get it only in a very distant future (by our estimate, in
$\sim 10$ billion years), when the phase transition is completed.

In Refs.~\ct{19,20,20a,21,22,22a} (and in our model) $a_Z$ plays the role
of the `acceleron', and a scalar boson $\sigma $, partner of the
acceleron, plays the role of the `inflaton' in the ``low scale
inflationary scenario" \ct{32}.

\subsection{Dark matter}

The existence of DM (non-luminous and non-absorbing matter) in the
universe is now well established. Candidates for non-baryonic DM
must be particles, which are stable on cosmological time scales.
They must interact very weakly with electromagnetic radiation.
Also they must have the right relic density. These candidates can
be black holes, axions, and weakly interacting massive particles
(WIMPs). In supersymmetric models WIMP candidates are the lightest
superparticles. The most known WIMP is the lightest neutralino.
WIMPs could be photino, higgsino, or bino.

In our model, as in Refs.~\ct{19,20,20a,21,22,22a}, mirror particles could
be considered as candidates of DM: mirror world interacts with
ordinary one only by gravity and really is a non-luminous and
non-absorbing dark matter. Fermions $\psi^{(Z)}_i$ introduced in
Subsection 5.1, could be considered as candidates for HDM (hot
dark matter), and their composites (`hadrons' of $SU(2)_Z$) could
play a role of the WIMPs in CDM. Investigating DM, it is possible
to search and study various signals such as: $\psi^{(Z)} + e \to
\psi^{(Z)} + e$, or $\psi^{(Z)} + N \to \psi^{(Z)} + N$, where $e$
is an electron and $N$ is a nucleon.

The detection of mirror particles: mirror quarks, leptons, Higgs
bosons, etc., could be performed at future colliders such as LHC.
Also the `messenger' scalar boson $\phi^{(Z)}$ can be produced at
LHC. Then the decay: $\phi_i^{(Z)} \to \ov{\psi}^{(Z)}_i + l$,
where $l$ stands for the SM lepton, can be investigated with
$\psi^{(Z)}$ as missing energies.

Leptogenesis and inflationary model also can be considered as
implications of our mirror physics. The full investigation is
beyond this paper.

\section{Conclusions and outlook}

We have discussed in this paper cosmological implications of the
parallel ordinary and mirror worlds in the case when the mirror
parity (MP) is not conserved. In our investigation the breakdown
of MP is characterized by the parameter $\zeta = v'/v$, where $v'$
and $v$ are the VEVs of the Higgs bosons -- Electroweak scales --
in the M- and O-worlds, respectively. During our numerical
calculations, we have used the values $\zeta = 10$ and $\zeta =
30$, in accordance with a cosmological estimate obtained in
Refs.~\ct{8,9,18}. We have shown that, as a result of the
MP-breaking, the evolutions of fine structure constants in O- and
M-worlds are not identical, and the extensions of the SM and SM$'$
are quite different if we want to have the same $E_6$-unification
in both worlds, what is predicted by Superstring theory \ct{7}. We
have assumed that the following chain of symmetry groups exists in
the ordinary world:
$$ SU(3)_C\times
SU(2)_L\times U(1)_Y\to
 SU(3)_C\times SU(2)_L \times SU(2)_R\times U(1)_X\times
 U(1)_Z$$
 $$\to SU(4)_C\times SU(2)_L \times SU(2)_R\times U(1)_Z\to
SO(10)\times U(1)_Z \to E_6\,.$$ Here we have chosen a  chain, which
leads to the asymptotically free $E_6$ unification, what is not
always fulfilled (see Ref.~\ct{7f}).

Then in the mirror M-world a simple logic dictates to consider the
following chain:
$$SU(3)'_C\times SU(2)'_L \times SU(2)'_Z  \times U(1)'_Y
$$ $$\to SU(3)'_C\times SU(2)'_L \times
SU(2)'_Z\times U(1)'_X\times
 U(1)'_Z$$ $$\to
SU(4)'_C\times SU(2)'_L\times SU(2)'_Z\times U(1)'_Z
$$
$$\to SU(6)'\times SU(2)'_Z \to E'_6\,.$$
The comparison of both evolutions in the ordinary and mirror
worlds is given in Figs.~5(a,b) and Figs.~6(a,b), where we have
presented the running of all fine structure constants. Here the SM
(SM$'$) is extended by MSSM (MSSM$'$), and then we see different
evolutions. Figs.~5(a,b) correspond to the SUSY breaking scales
$M_{SUSY}=1$ TeV and $M'_{SUSY}=10$ TeV, while Figs.~6(a,b) are
presented for $M_{SUSY}=10$ TeV and $M'_{SUSY}=300$ TeV, according
to the MP-breaking parameters $\zeta = 10$ and $\zeta = 30$,
respectively. We see $M_{SGUT}\approx 2.4\cdot 10^{17}$ GeV and
$\alpha_{SGUT}^{-1}\approx 26.06$ -- for Figs.~5(a,b), and
$M_{SGUT}\approx 6.96\cdot 10^{17}$ GeV and
$\alpha_{SGUT}^{-1}\approx 27.64$ -- for Figs.~6(a,b). It is quite
significant to emphasize that in these cases mirror right-handed
neutrinos appear only at the scale $\sim 10^{17}$ GeV, close to
the $E'_6$ unification.

The (super)grand unification $E'_6$ is based on the group
$E'_6\supset SU(6)'\times SU(2)'_Z$, and the presence of a new
unbroken gauge group $SU(2)'_Z$ in the mirror world gives
significant consequences for cosmology: it explains the
`quintessence' model of our accelerating universe (see
Refs.~\ct{19,20,20a,21,22,22a}). As we have shown in Figs.~3(a,b) and
4(a,b) for the mirror world (and also in Figs.~5(a,b) and 6(a,b)
for both worlds), the $SU(2)'_Z$ gauge coupling, presented by
${\alpha'}_{2Z}^{-1}(\mu)$, takes its initial value at the
superGUT scale $\sim 10^{18}$ GeV: ${\alpha'}_{2Z}^{-1}(M'_{SGUT})
= \alpha^{-1}_{E6}$, and then runs down to very low energies,
leading to the extremely strong coupling constant at the scale
$\Lambda_Z\sim 10^{-3}$ eV. At this scale $SU(2)'_Z$ instantons
induce a potential for an axion-like scalar particle $a_Z$, which
can be called `acceleron', because it gives the value $w=-1$ and
leads to the acceleration of our universe. The existence of the
scale $\Lambda_Z\sim 10^{-3}$ eV explains the value of
cosmological constant: $CC\approx (3\times 10^{-3}\,\,\rm{
eV})^4$, which is given by cosmological measurements
\ct{24,25,25a,26,27,27a,28}.

It was assumed in Refs.~\ct{20,20a,21,22,22a} that at present time our
universe exists in the `false' vacuum given by the axion
potential. The universe will live there for a long time and its $CC$
(measured in cosmology) is tiny, but nonzero. However, at the end
our universe will jump into the `true' vacuum and will get a zero
$CC$. But this problem is not trivially solved, and at present we
have only a hypothesis.

Now it is worth the reader's attention to observe that in the
mirror world we have three scales (presumably corresponding to the
three MPP vacua \ct{35}): $$\Lambda_1 = \Lambda_Z\sim
10^{-12}\,\,\rm {GeV}, \quad \Lambda_2 = \Lambda_{EW}\sim 10^3
\,\,\rm{GeV}\quad\rm{ and}\quad  \Lambda_3 = \Lambda_{SGUT}\sim
10^{18}\,\, \rm {GeV}\,.$$ It is not difficult to notice that they
obey the following interesting relation:
$$\Lambda_1\cdot \Lambda_3 \approx \Lambda_2^2\,.$$
In our model of the universe with broken mirror parity we have the
following particle content of the group $SU(2)'_Z$, which does not
coincide with Refs.~\ct{19,20,20a,21,22,22a}:

\begin{itemize}

\item two doublets of fermions $\psi^{(Z)}_i$ ($i = 1,2$),
         or a triplet of fermions $\psi^{(Z)}_f$  ($f = 1,2,3$);

\item two doublets of scalar fields $\phi^{(Z)}_i$ ($i =
1,2$).

\end{itemize}

Also as in Refs.~\ct{19,20,20a,21,22,22a}, we have considered a complex
singlet scalar field $\varphi_Z$, which produces `acceleron' $a_Z$
and `inflaton' $\sigma_Z$ and gives an analogous `quintessence'
model of our universe.

Unfortunately, we cannot predict exactly the scales $M_{SUSY}$ and
$M_R$ presented in our Figs.~1-6. The quantitative description of
the model depends on these scales. Nevertheless, we hope that a
qualitative scenario for the evolution of our universe, given in
Refs.~\ct{19,20,20a,21,22,22a} and in the present paper, is valid.

In Section 6 we have discussed a possibility to consider the
fermions $\psi^{(Z)}_i$ as candidates for the WIMP CDM. Searching
DM, it is possible to observe and study various signals of these
particles.

Also it is quite significant that this investigation opens the
possibility to fix a grand unification group ($E_6$ ?)  from
cosmology.

\section*{Acknowledgments}

C.R.D. deeply thanks Prof.~R.N.Mohapatra for useful advices.

This work was supported by the Russian Foundation for Basic
Research (RFBR), project No 05-02-17642.

\clearpage\newpage \bfi \centering
\includegraphics[height=100mm,keepaspectratio=true,angle=0]{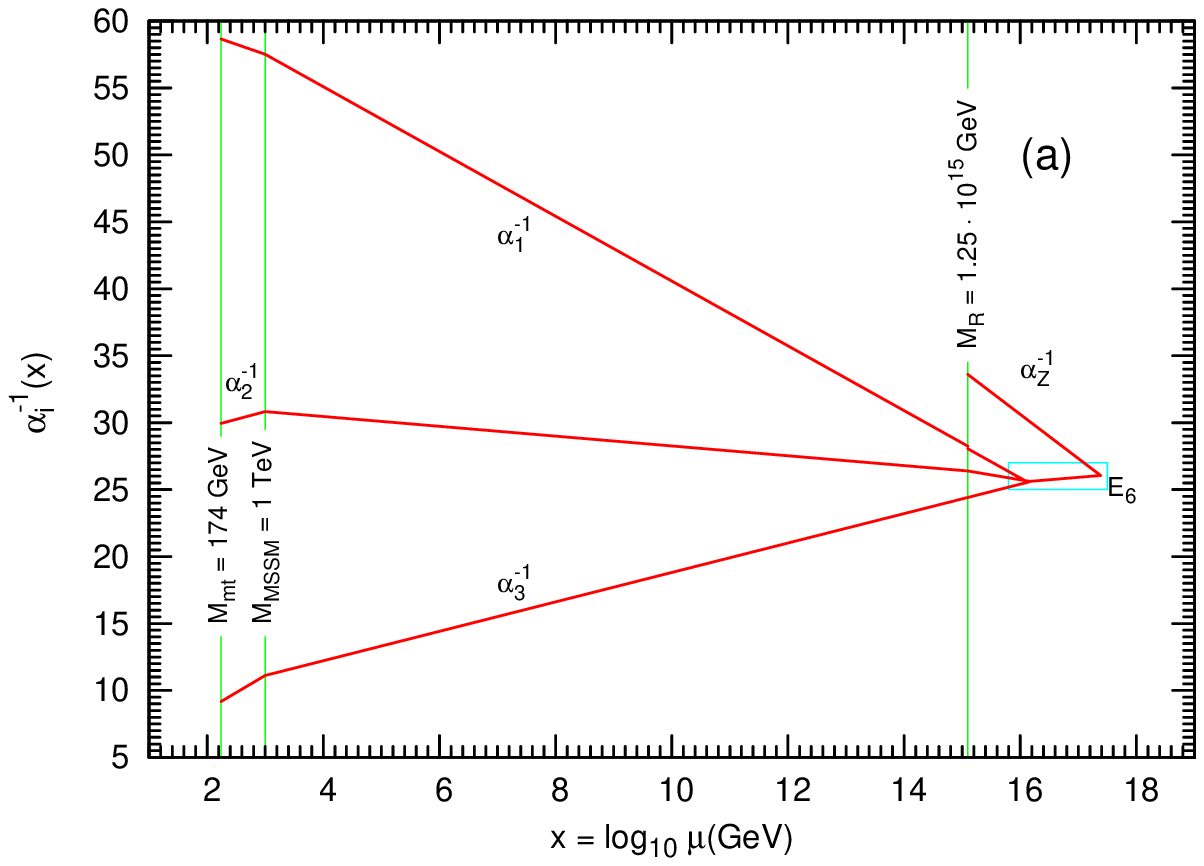}
\includegraphics[height=100mm,keepaspectratio=true,angle=0]{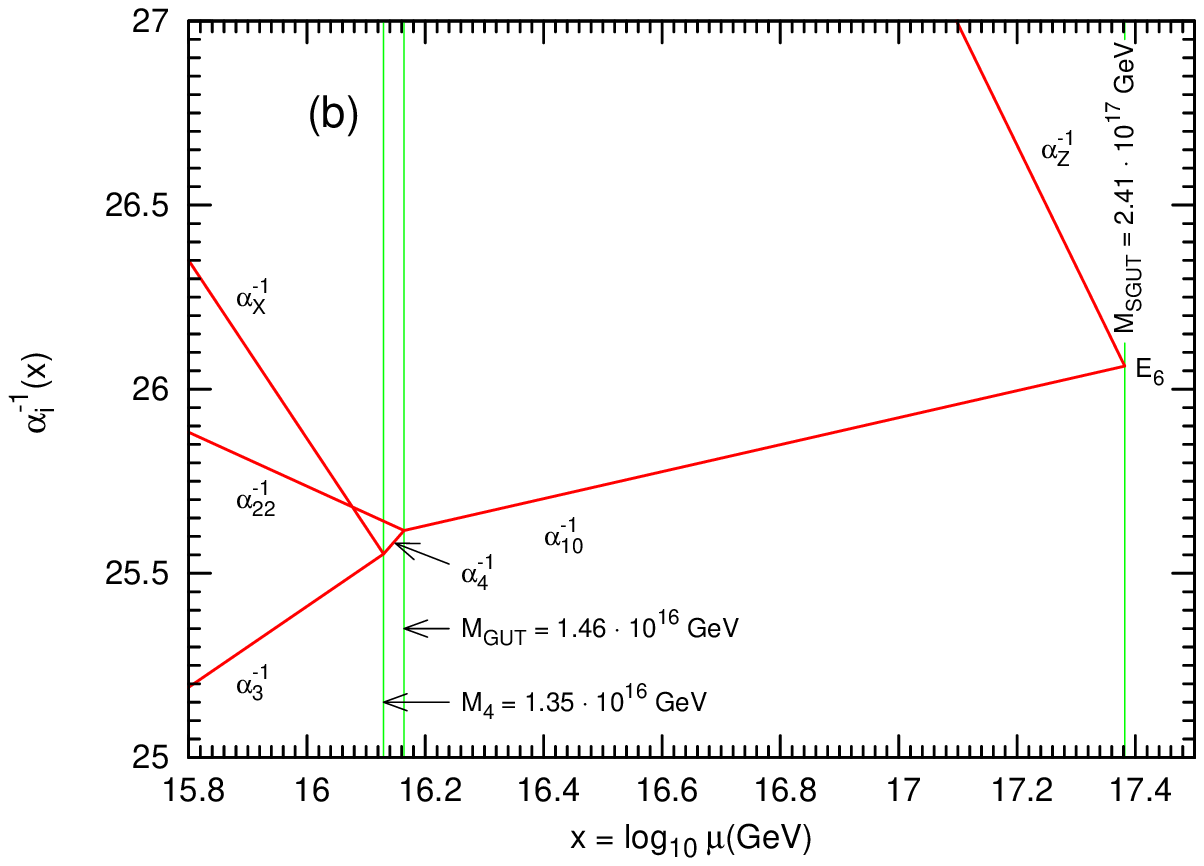}
\caption { Figure (a) presents the running of the
inverse coupling constants $\alpha_i^{-1}(x)$ in the ordinary
world from the Standard Model up to the $E_6$ unification for SUSY
breaking scale $M_{SUSY}= 1$ TeV and seesaw scale $M_R=1.25\cdot
10^{15}$ GeV. This case gives: $M_{SGUT}\approx 2.4\cdot 10^{17}$
GeV and $\alpha_{SGUT}^{-1}\approx 26.06$. (b) is same as (a),
but zoomed in the scale region $10^{15.8}$ GeV up  to the $E_6$
unification to show the details.}\efi

\clearpage\newpage \bfi \centering
\includegraphics[height=100mm,keepaspectratio=true,angle=0]{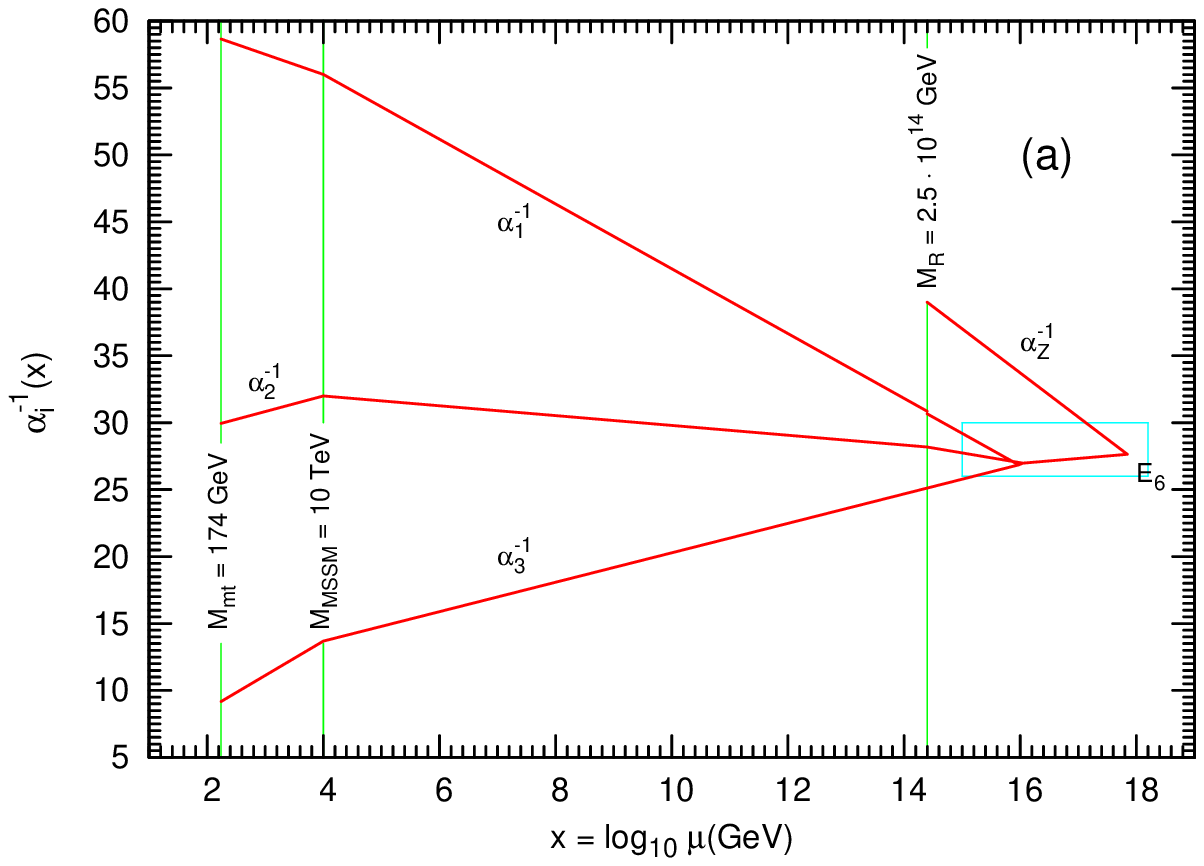}
\includegraphics[height=100mm,keepaspectratio=true,angle=0]{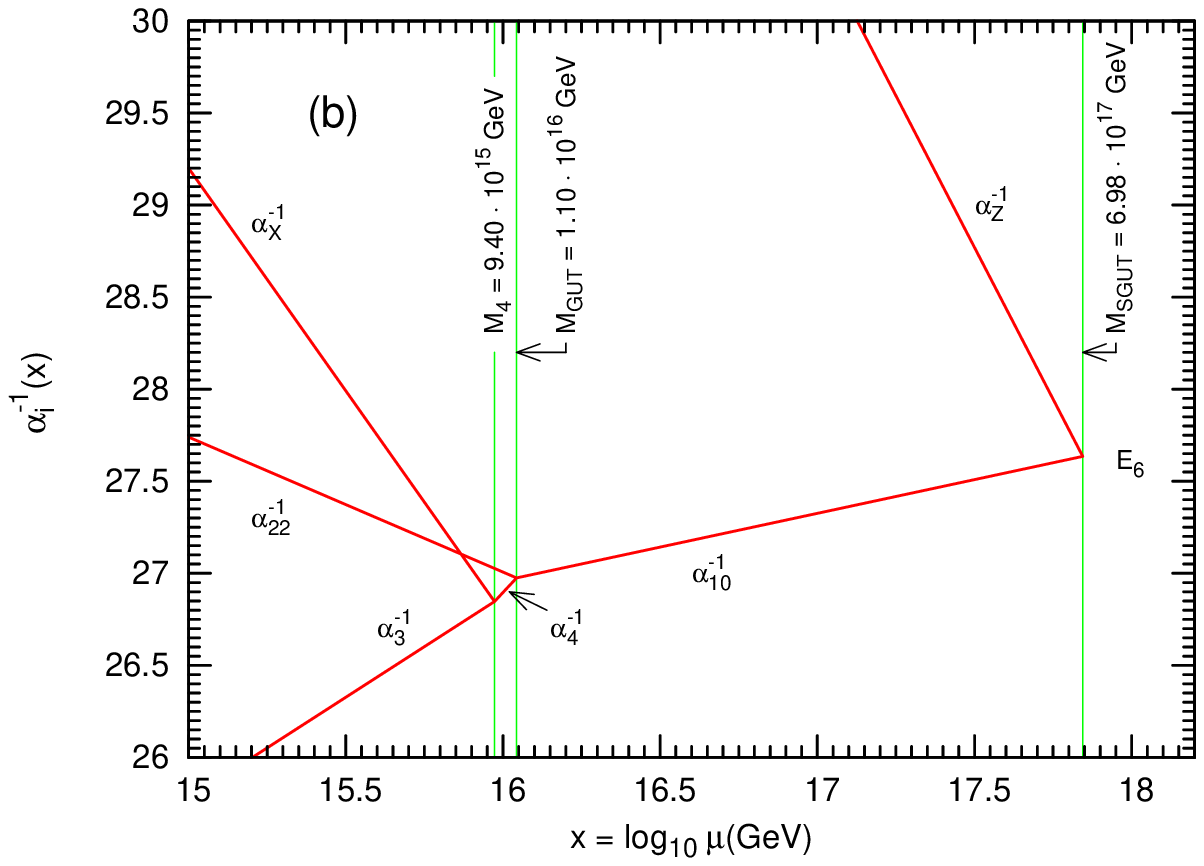}
\caption {Figure (a) presents the running of the
inverse coupling constants $\alpha_i^{-1}(x)$ in the ordinary
world from the Standard Model up to the $E_6$ unification for SUSY
breaking scale $M_{SUSY}= 10$ TeV and seesaw scale $M_R=2.5\cdot
10^{14}$ GeV. This case gives: $M_{SGUT}\approx 6.96\cdot 10^{17}$
GeV and $\alpha_{SGUT}^{-1}\approx 27.64$. (b) is same as (a),
but zoomed in the scale region $10^{15}$ GeV up  to the $E_6$
unification to show the details.}\efi

\clearpage\newpage \bfi \centering
\includegraphics[height=100mm,keepaspectratio=true,angle=0]{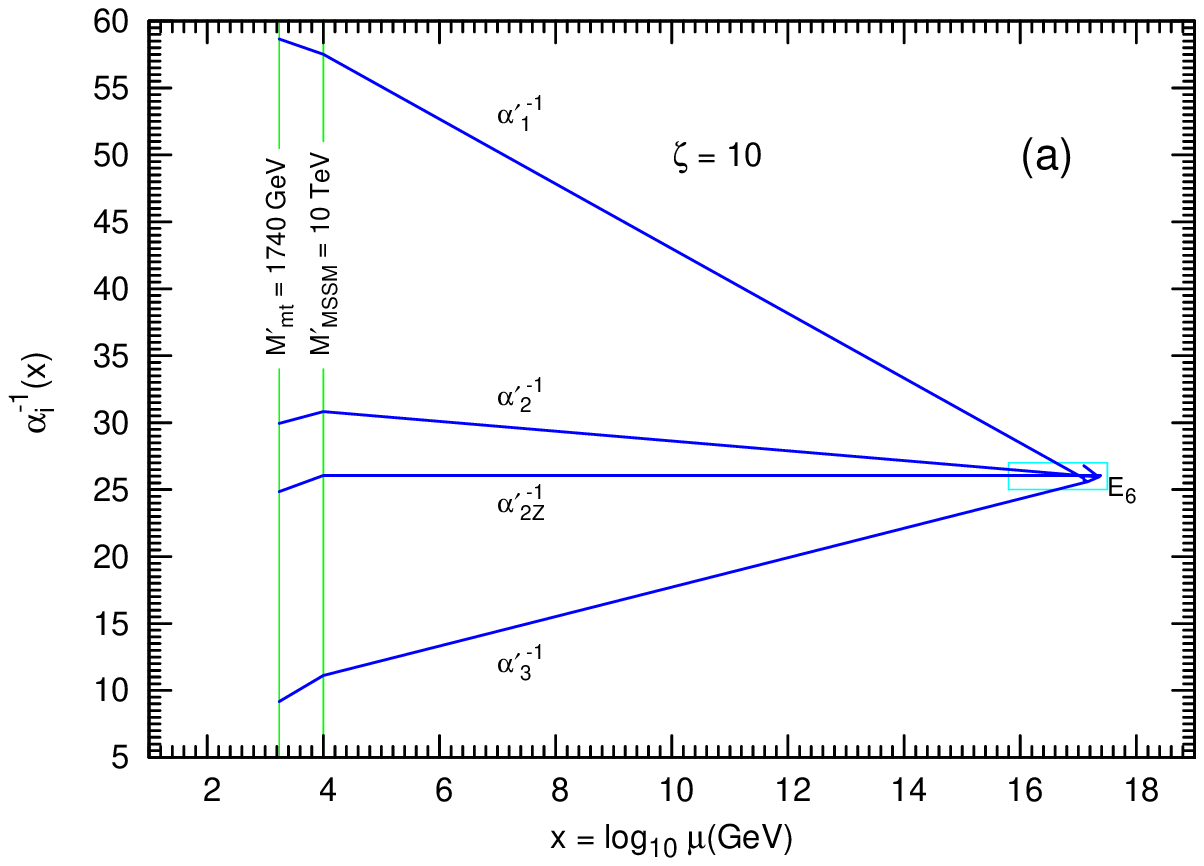}
\includegraphics[height=100mm,keepaspectratio=true,angle=0]{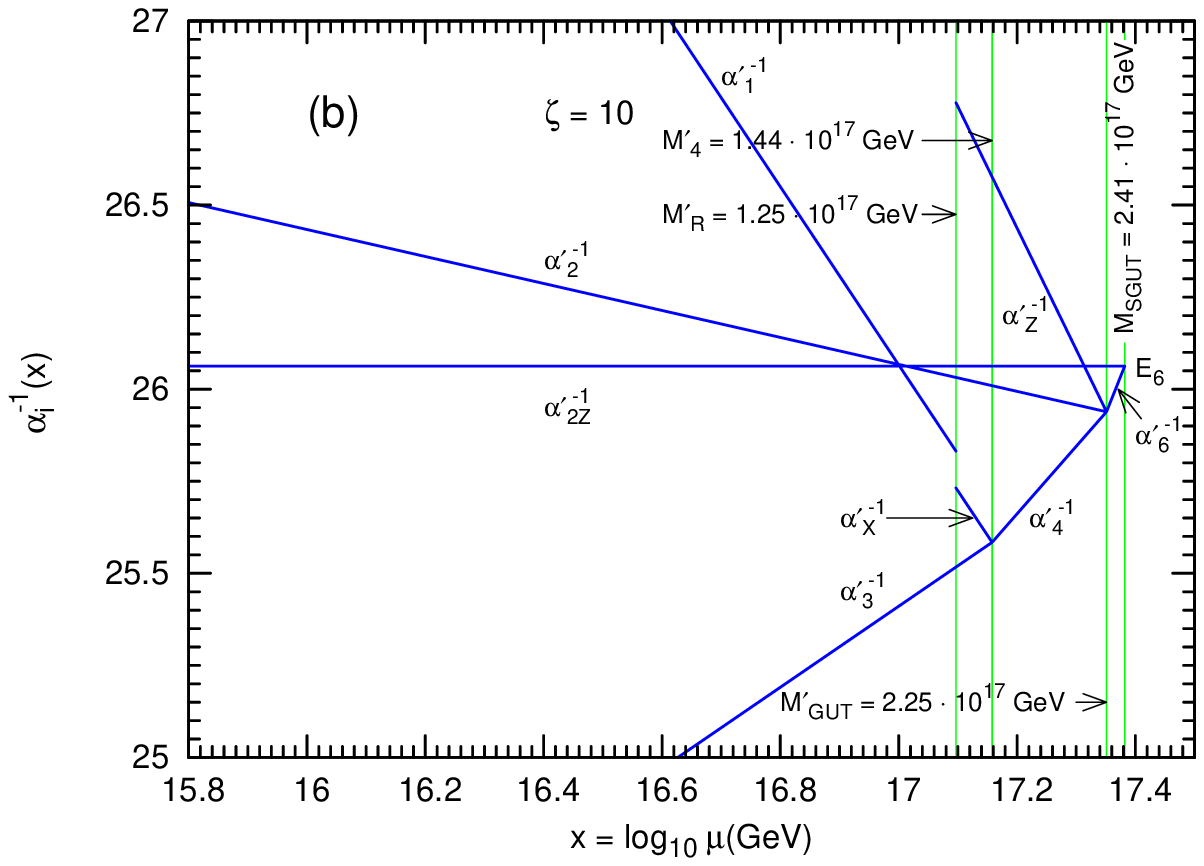}
\caption {Figure (a) presents the running of the
inverse coupling constants $\alpha_i^{-1}(x)$ in the mirror world
from the Standard Model up to the $E_6$ unification for SUSY
breaking scale $M'_{SUSY}= 10$ TeV and mirror seesaw scale
$M'_R=1.44\cdot 10^{17}$ GeV; $\zeta = 10$. This case gives:
$M_{SGUT}\approx 2.4\cdot 10^{17}$ GeV and
$\alpha_{SGUT}^{-1}\approx 26.06$. (b) is same as (a),
but zoomed in the scale region $10^{15.8}$ GeV up  to the $E_6$
unification to show the details.} \efi

\clearpage\newpage \bfi \centering
\includegraphics[height=100mm,keepaspectratio=true,angle=0]{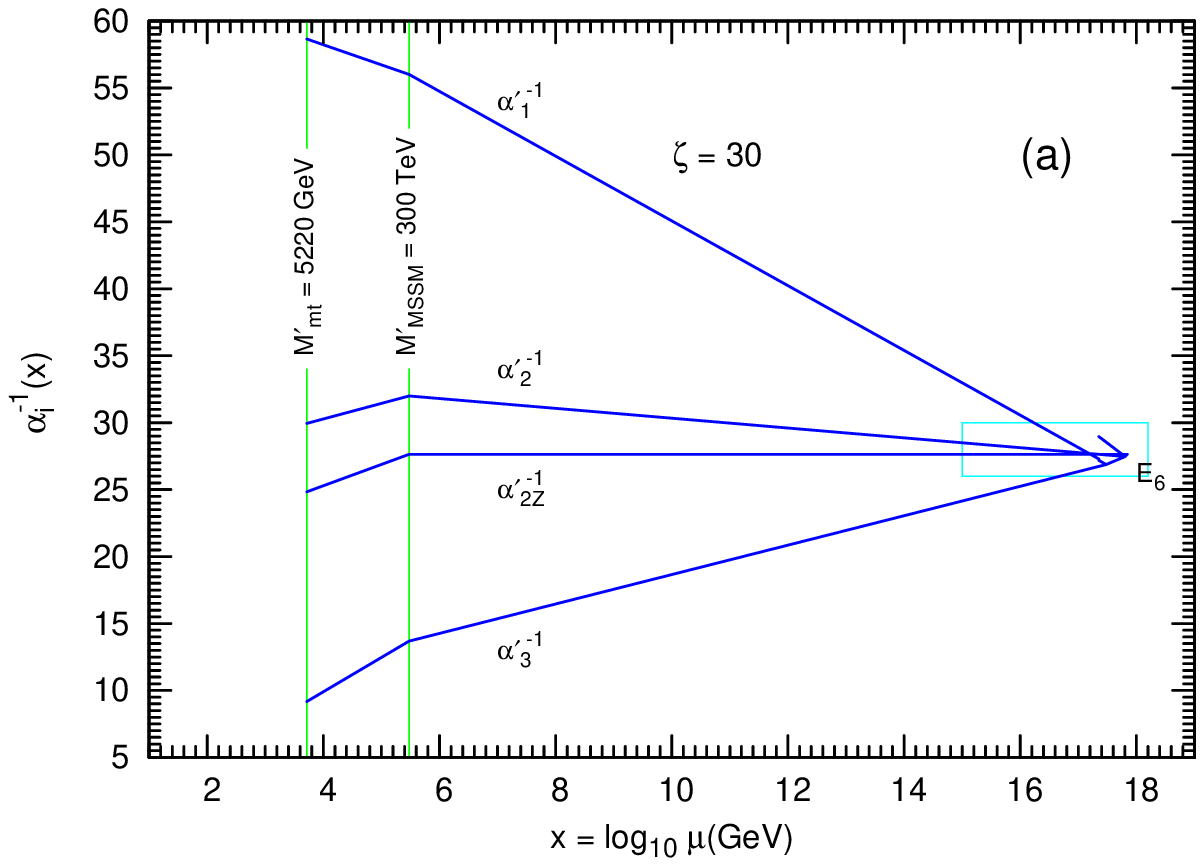}
\includegraphics[height=100mm,keepaspectratio=true,angle=0]{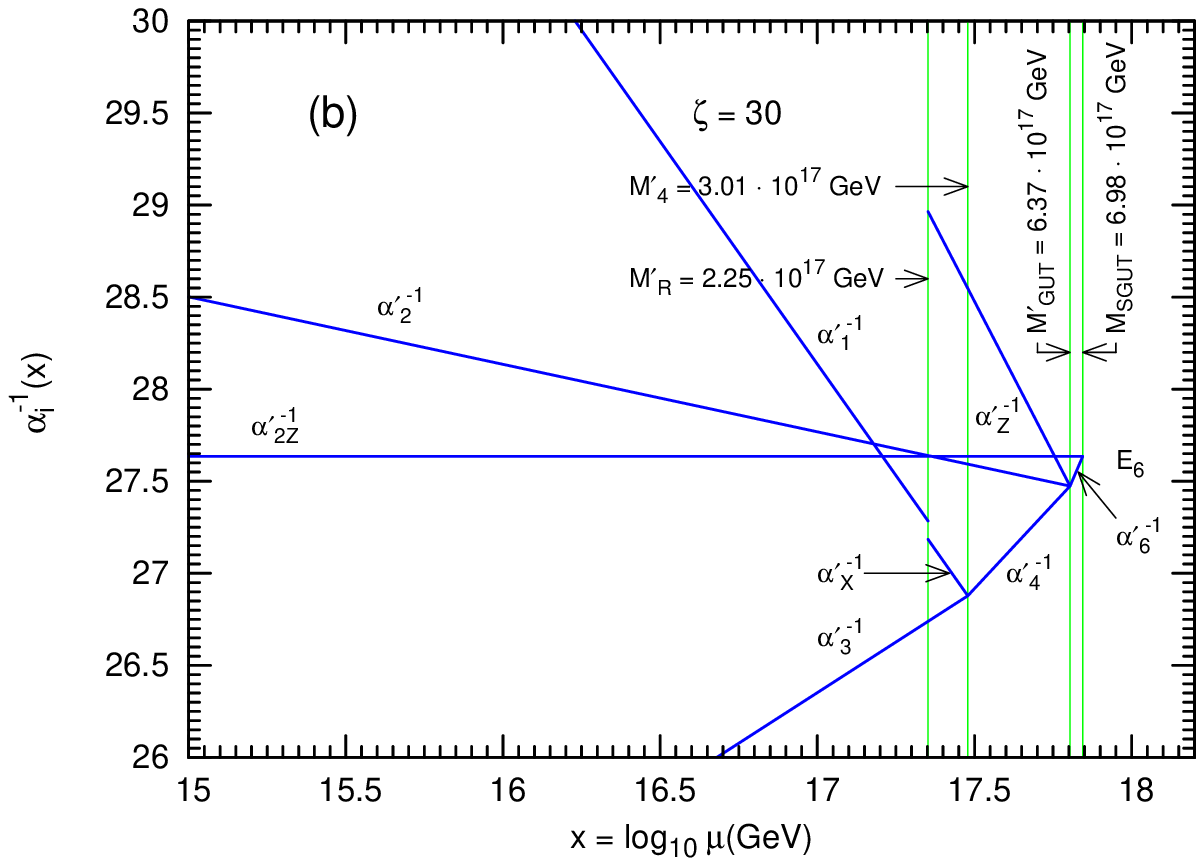}
\caption {Figure (a) presents the running of the
inverse coupling constants $\alpha_i^{-1}(x)$ in the mirror world
from the Standard Model up to the $E_6$ unification for SUSY
breaking scale $M'_{SUSY}= 300$ TeV and mirror seesaw scale
$M'_R=2.25\cdot 10^{17}$ GeV; $\zeta = 30$. This case gives:
$M_{SGUT}\approx 6.96\cdot 10^{17}$ GeV and
$\alpha_{SGUT}^{-1}\approx 27.64$. (b) is same as (a),
but zoomed in the scale region $10^{15}$ GeV up  to the $E_6$
unification to show the details.} \efi

\clearpage\newpage \bfi \centering
\includegraphics[height=100mm,keepaspectratio=true,angle=0]{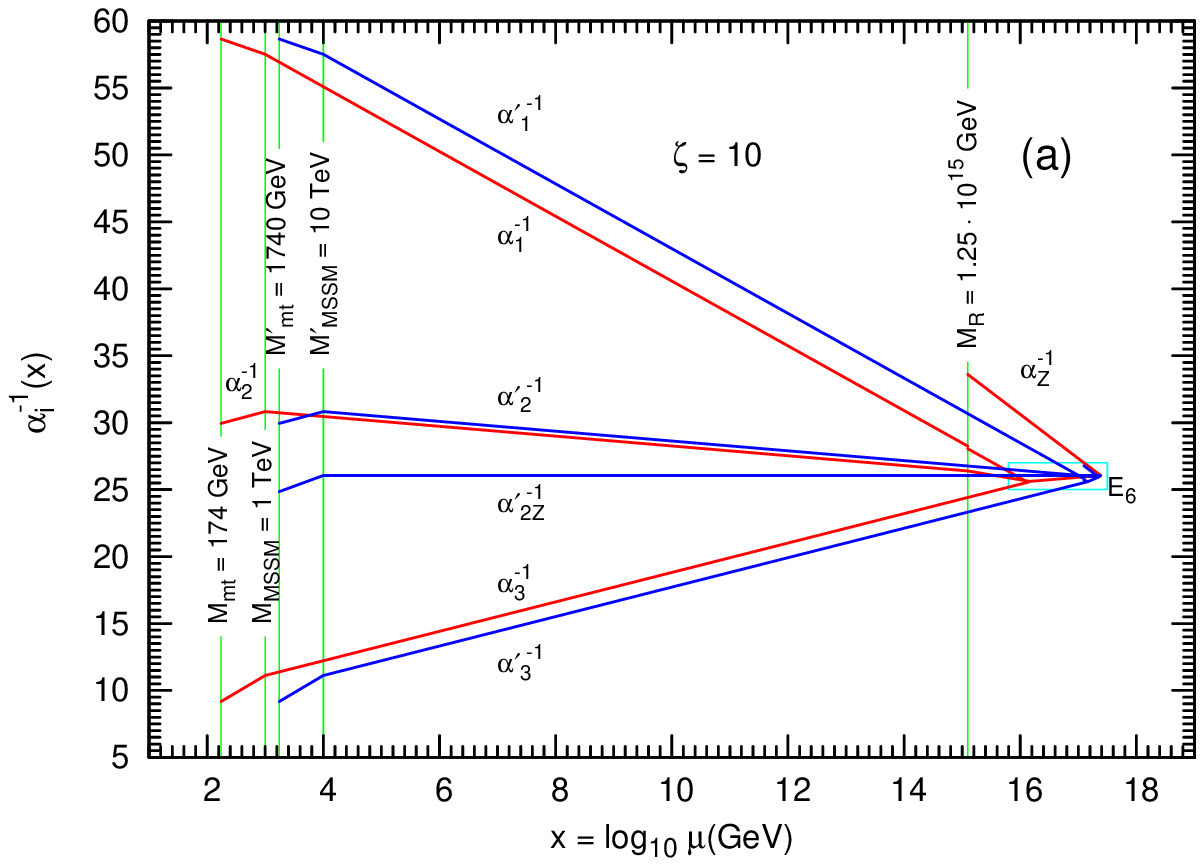}
\includegraphics[height=100mm,keepaspectratio=true,angle=0]{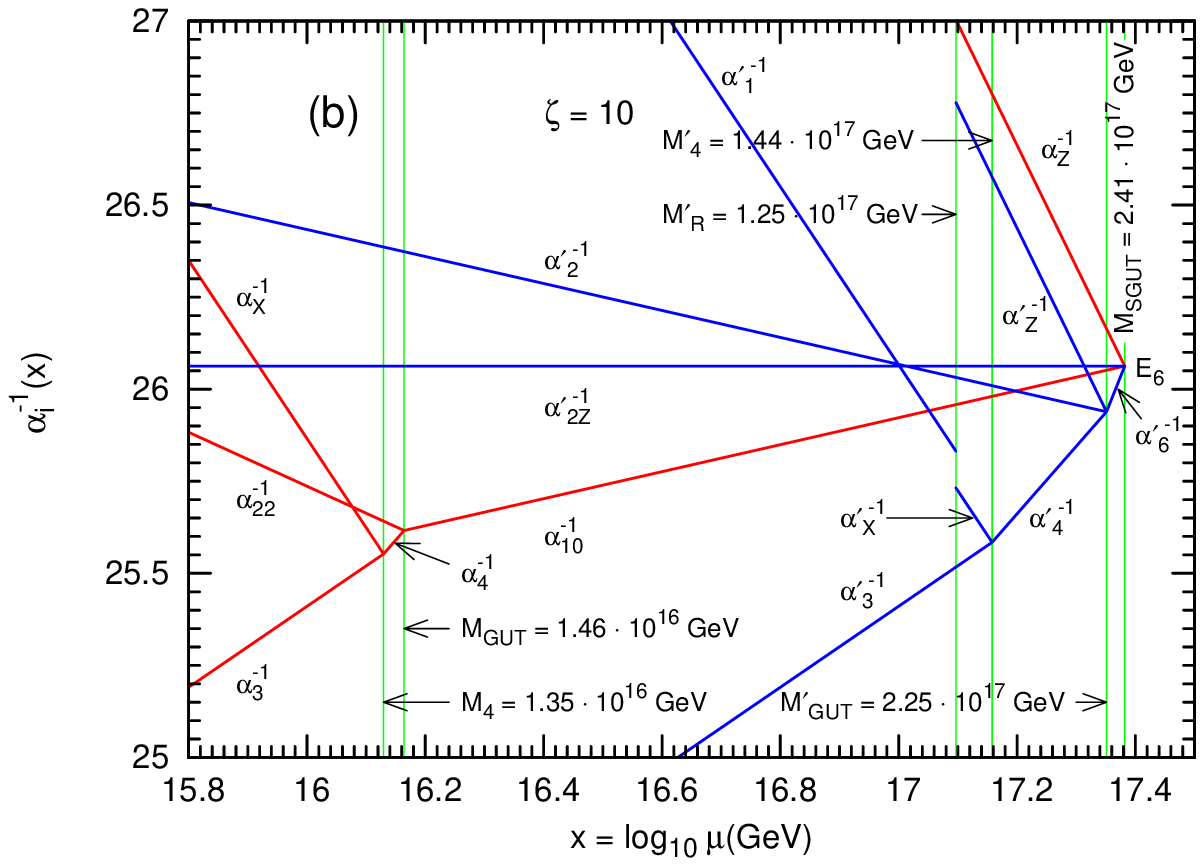}
\caption {In (a) the running of the inverse coupling
constants $\alpha_i^{-1}(x)$ in both ordinary and mirror worlds
with broken mirror parity from the Standard Model up to the $E_6$
unification for SUSY breaking scales $M_{SUSY}= 1$ TeV,
$M'_{SUSY}= 10$ TeV and seesaw scales $M_R=1.25\cdot 10^{15}$ GeV,
$M'_R=1.44\cdot 10^{17}$ GeV; $\zeta = 10$. This case gives:
$M_{SGUT}\approx 2.4\cdot 10^{17}$ GeV and
$\alpha_{SGUT}^{-1}\approx 26.06$. (b) is same as (a),
but zoomed in the scale region $10^{15.8}$ GeV up  to the $E_6$
unification to show the details.} \efi

\clearpage\newpage \bfi \centering
\includegraphics[height=100mm,keepaspectratio=true,angle=0]{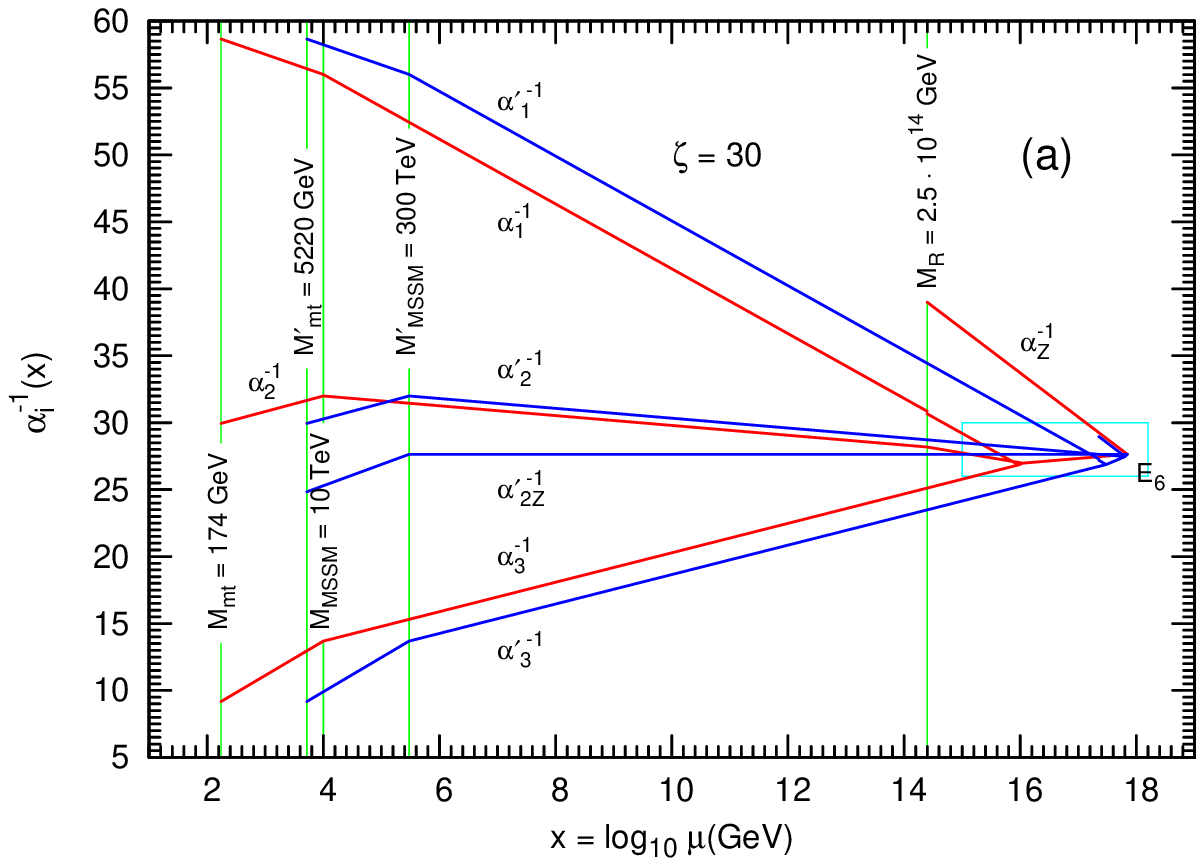}
\includegraphics[height=100mm,keepaspectratio=true,angle=0]{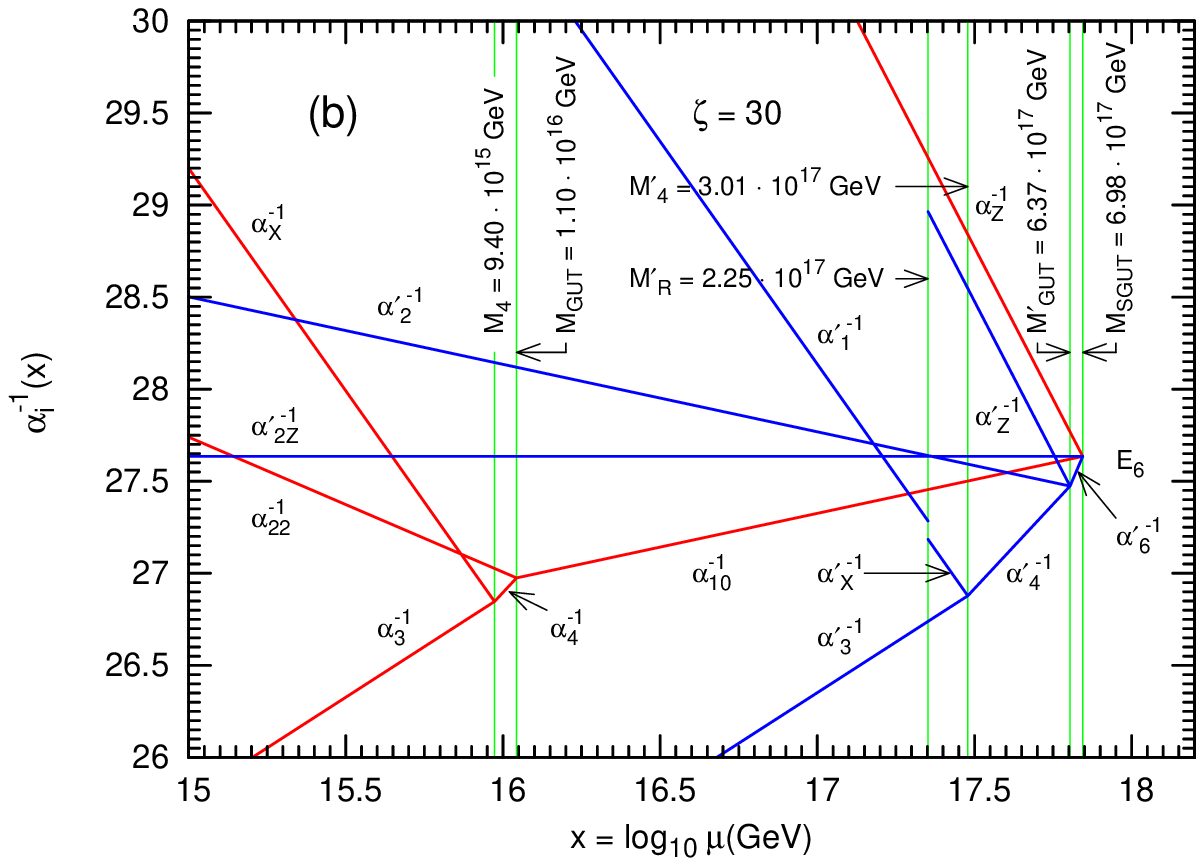}
\caption {In (a) the running of the inverse coupling
constants $\alpha_i^{-1}(x)$ in both ordinary and mirror worlds
with broken mirror parity from the Standard Model up to the $E_6$
unification for SUSY breaking scales $M_{SUSY}= 10$ TeV,
$M'_{SUSY}= 300$ TeV and seesaw scales $M_R=2.5\cdot 10^{14}$ GeV,
$M'_R=2.25\cdot 10^{17}$ GeV; $\zeta = 30$. This case gives:
$M_{SGUT}\approx 6.96\cdot 10^{17}$ GeV and
$\alpha_{SGUT}^{-1}\approx 27.64$. (b) is same as (a),
but zoomed in the scale region $10^{15}$ GeV up  to the $E_6$
unification to show the details.} \efi

\clearpage\newpage \bfi \centering
\includegraphics[height=100mm,keepaspectratio=true,angle=0]{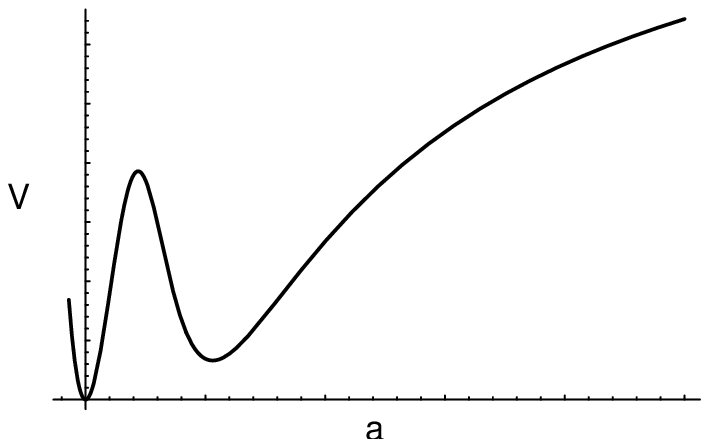}
\caption {The axion potential $V$ as a function of $ a=|a_Z|$.
It shows the `true' vacuum at $ \langle a_Z\rangle = 0$ and the `false' vacuum
at $\langle a_Z\rangle = 2\pi v_Z$.} \efi

\end{document}